\documentstyle[11pt]{article}

\newcommand{\half}{{\scriptstyle{\frac{1}{2}}}}

\newcommand{\dl}{\delta^{(3)}({\bf r})}

\newcommand{\BE}{\begin{equation}}
\newcommand{\EE}{\end{equation}}
\newcommand{\BA}{\begin{eqnarray}}
\newcommand{\EA}{\end{eqnarray}}
\newcommand{\vol}{{\sf V}}
\newcommand{\num}{{\sf N}}

\begin{document}
\begin{titlepage}

\vspace*{1mm}
\begin{center}

            {\LARGE{\bf Approximate Lorentz invariance of 
             the vacuum:\\ a physical solution of the `hierarchy problem' ? }}

\vspace*{14mm}
{\Large  M. Consoli }
\vspace*{4mm}\\
{\large
Istituto Nazionale di Fisica Nucleare, Sezione di Catania \\
c/o Dipartimento di Fisica, Citt\`a Universitaria \\
Via Santa Sofia 64, 95123 Catania, Italy}
%Corso Italia 57, 95129 Catania, Italy}
%\vspace*{3mm}\\
%and \\
%\vspace*{3mm}
%{\Large F. Siringo}
%\vspace*{2.5mm}\\
%{\large Dipartimento di Fisica dell' Universit\`a di Catania \\
%Corso Italia 57, 95129 Catania, Italy}
%\vspace{7mm}\\
\end{center}
\begin{center}
{\bf Abstract}
\end{center}

In the `condensed phase' of
effective quantum field theories one expects
deviations from exact Lorentz
invariance at ultralow momenta $|{\bf{k}}| < \delta$ where the shell $\delta$ 
should only vanish
in the strict local limit of the theory when the ultraviolet cutoff 
$\Lambda \to \infty$.
I explore this idea for the Higgs condensate suggesting that, in this case, 
the resulting relation connecting $\delta$, $\Lambda$ and the Fermi scale 
might provide a simple physical solution of the `hierarchy problem'.
In this picture, the Planck scale is not a
purely ultraviolet quantity but embodies in its 
numerical value the peculiar 
infrared-ultraviolet connection that is realized in the
scalar condensate.
\vskip 35 pt
\end{titlepage}

\section{Introduction}

{\bf 1.1} Following analogies with condensed matter, 
the idea of a `condensed vacuum' is now playing a very important role in
modern particle physics. Indeed, in many different contexts one introduces
a set of elementary quanta whose perturbative `empty' vacuum state 
$|o\rangle$ is not the physical ground state of the interacting theory. 
In the physically relevant case of the Standard Model,
the situation can be summarized saying \cite{thooft} that 
 "What we experience as 
empty space is nothing but the configuration of the Higgs field that has the
lowest possible energy. If we move from field jargon to particle jargon, this 
means that empty space is actually filled with Higgs particles. They have 
Bose condensed." 

This type of conclusion is also favoured by
the experimental observation that
bodies can flow in the vacuum without any apparent friction. This leads to
the physical picture of a superfluid, a result that can easily be
 understood in the Standard Model where the restoring
critical temperature is so high. Thus,  
for all practical purposes, 
the scalar condensate might be considered  
a zero-temperature Bose liquid.

Clearly, this type of medium is not
the ether of classical physics. However, it is also different from
the `empty' space-time of Special Relativity, assumed at the base of 
axiomatic quantum field theory. 
Thus, it is not unconceivable that
the macroscopic occupation of the same quantum
state can represent, in some appropriate sense, the operative construction
of a preferred frame, a `quantum ether'. This might account for
that particular type of non-locality which is required
by a `realistic' description of EPR experiments 
\cite{hardy,scarani} and by
 the observed violations of Bell's inequalities \cite{tittel}.

In connection with the idea of ether, it should be better 
underlined, perhaps, that the original Einstein's point of view 
had been later reconsidered with the transition 
from Special Relativity to General Relativity. Probably, he realized 
that Riemannian geometry (with its covariant 
derivatives, its Christoffel symbols,...) is also the natural
framework to describe the dynamics of deformable media
(see the appendix
of ref.\cite{sommerfeld}). In fact, 
for this or other reasons, in 1919, in the
`Morgan Manuscript' \cite{barone,kostro1}, he wrote "..in 1905 I was of the
opinion that it was no longer allowed to speak about the ether in physics. 
This opinion, however, was too radical as we will see later when we 
consider the general relativity theory. It is allowed much more than 
before to accept a medium penetrating the whole space and to regard the
electromagnetic fields and the matter as well as states of it. But it is 
not allowed to attribute to this medium, in analogy to ponderable matter, 
a state of motion in any point. This ether must not be conceived as 
composed of particles the identity of which can be followed in time...
One can thus say that the ether is resurrected in the general 
theory of relativity, though in a more sublimated form." 

This `resurrection' of ether in Einstein's
mind is confirmed by a large amount of published and
unpublished manuscripts that have now been collected in a book 
by Kostro \cite{kostro2}.
I just observe that, in a picture of the vacuum as a quantum Bose liquid, the
last statements would be easy to understand. In fact, the properties of such 
a medium depend in an essential way 
on the quantum nature of its constituents. This requires 
a form of quantum hydrodynamics, of the type originally considered
by Landau \cite{hydrolandau}, where the local density of the fluid 
$n({\bf{r}})$ and 
the current density vector ${\bf{J}}({\bf{r}})$ 
have canonical commutation relations 
 as in quantum mechanics for the position
and momentum operators. 

Therefore, following the original evolution of Einstein's thought, one may
attempt to introduce an underlying quantum ether in connection with gravity 
by choosing the particle physics vacuum as
the most natural candidate. To this end, however, we have to 
start to consider the Higgs condensate as a real physical medium 
and improve on the simplest approximation where it is treated as a purely 
classical c-number field. 

To better appreciate the potentiality of 
this approach, I observe that the idea of generating 
gravity from the excitations of a `self-sustaining' superfluid 
is particularly appealing since it
leads to the conclusion that only its
 {\it departures from equilibrium} contribute 
to the space-time curvature \cite{volo2}. This provides a simple physical 
argument to understand 
why the huge energy density of the {\it unperturbed}
vacuum plays no role and, in this sense, 
the gravity-superfluid connection follows naturally from
Feynman's indication "...the first thing we should 
understand is how to formulate gravity so that it doesn't interact with the 
energy in the vacuum" \cite{rule}.

The analogies between gravity and the density fluctuations of some
physical medium have been 
explored by several authors in different frameworks (for a
 comprehensive list see ref.\cite{barcelo}). I just observe that
a model of gravity from an underlying scalar
ether has been proposed in ref.\cite{arminjon}. In a different context, 
an acoustic analogy of gravity
has been constructed in refs.\cite{unruh,visser,liberati} when studying
the propagation of density
fluctuations in a moving fluid. This requires the introduction of 
an `acoustic metric' modelled on
Galilei covariance, and leads to
the acoustic equivalent concepts of black holes, Hawking radiation,...
In this framework, the analog of gravitational black holes, as 
structures that can be formed in dilute Bose-Einstein condensates, 
has also been explicitely considered \cite{garay}.

However, if the density fluctuations of the physical vacuum are really
governed by a Galilei-covariant
 acoustic metric, why Lorentz covariance works so well ? 
Equivalently, why gravity is so weak in ordinary experimental conditions 
as compared to the other effects (say electromagnetism) for which an exact
Lorentz-covariant description is possible ?

As a possible answer to these questions, 
one can exploit the analogy with an infinite isotropic elastic medium. 
Its infinitesimal deformation 
at a given point ${\bf{r}}$ at the time $t$, say
${\bf{Z}}({\bf{r}},t)$, is governed by the source-free
partial differential equation
\cite{sommerfeld,elalandau}
\BE
\label{ela1}
(c^2_t \Delta - {{\partial^2}\over{\partial t^2}}) {\bf{Z}} + 
{{c^2_t}\over{1-2\nu}} {\bf{\nabla}}(\nabla\cdot {\bf{Z}})=0
\EE
In the above equation, I have introduced the square velocity 
$c^2_t\equiv {{Y}\over{2\rho (1+\nu)}}$, where 
$\rho$ denotes the density of the medium, 
$Y$ the Young modulus and $\nu$ the Poisson ratio
($0\leq  \nu \leq 1/2$ 
with $\nu\to 1/2$ defining the incompressibility limit). 

Now, by expressing ${\bf{Z}}={\bf{S}}+{\bf{T}}$ such that 
$\nabla$x${\bf{S}}=0$ and $\nabla\cdot {\bf{T}}=0$, and introducing 
$\tau \equiv c_t t$, Eq.(\ref{ela1}) implies
the propagation of two type of waves: transverse waves of distortion
\BE
\label{ela2}
(\Delta - {{\partial^2}\over{\partial \tau^2}}) {\bf{T}}=0
\EE
and longitudinal waves of dilatation 
\BE
\label{ela3}
(\eta \Delta - {{\partial^2}\over{\partial \tau^2}}) {\bf{S}}=0
\EE
whose velocity is 
\BE
\label{ela4}
c^2_s=\eta c^2_t > c^2_t
\EE
with 
\BE
\eta= (1 +{{1}\over{1-2\nu}}) > 1
\EE
For any value of $\nu$ 
no linear transformation $({\bf{r}},\tau) \to ({\bf{r}}',\tau')$ can preserve
the form-invariance of both the differential operators in
 Eqs.(\ref{ela2}) and (\ref{ela3}). In this sense,
there are two separate forms of `Lorentz-covariance' (associated with $c=c_s$ 
or $c=c_t$) and no unified description is possible.

The physically relevant situation arises in the limit $\nu \to 1/2$ where 
$\eta \to \infty$. In this case, 
$c_s$-Lorentz-covariance reduces to Galilei covariance. 
Further, any observer with speed 
$|{\bf{v}}| \ll c_t \ll c_s$ sees no appreciable 
Doppler shift of the frequencies 
or of the wave vectors of the longitudinal waves. Indeed, with 
respect to the corresponding effect for the transverse waves,
 these are suppressed by the very small number $c_t/c_s\to 0$. Finally, 
under the action of an external force ${\bf{f}}\equiv \nabla Q$, 
Eq.(\ref{ela3}) becomes
\BE
\label{ela5}
(\eta \Delta - {{\partial^2}\over{\partial \tau^2}}) {\bf{S}}=\nabla Q
\EE
so that, by defining ${\bf{S}}=\nabla \varphi$, one gets 
\BE
\label{ela6}
         \Delta \varphi({\bf{r}},t) = {{1}\over{\eta}}Q({\bf{r}},t)
\EE
up to infinitesimal higher order $1/\eta^2$ 
corrections. Thus, the longitudinal
waves of dilatation are seen as an instantaneous effect and
the presence of the huge velocity 
$c_s $ remains
`hidden' in the suppression factor $1/\eta$ of the strength of $Q$. 

In this paper, starting from a microscopic
description of the scalar condensate, I'll construct the equivalent
of the $c_s/c_t \to \infty$ limit, with $c_t=c$, thus 
obtaining a possible solution of the so called
`hierarchy problem' between the Fermi scale
of electroweak interaction and the Planck scale. 
In this sense, it turns out that gravity can be {\it induced} by the phenomenon
of spontaneous symmetry breaking. Some
peculiar aspects of this idea, however, have no obvious counterpart in the
traditional `induced-gravity' approach \cite{fuji,zee,adler}.
For this reason, I'll suggest in the
end a possible way to relate the two descriptions.

\vskip 10 pt
{\bf 1.2} For my analysis, I'll start from scratch, i.e. from
the more conventional 
point of view that the phenomenon of vacuum condensation is
just a convenient way to rearrange the set of original degrees of freedom and
can peacefully coexist with Lorentz invariance. Actually, some authors 
\cite{bennet} realized that a non-trivial vacuum
structure has to imply some form of non-locality but
concluded that it should be possible to reabsorb all differences between
the physical vacuum and empty space into 
some basic parameters such as the particle masses and few physical
constants. 

However, on a general ground, the coexistence of 
{\it exact} Lorentz invariance and vacuum condensation in {\it effective}
quantum field theories is not so trivial. In fact, as a consequence of the
violations of locality at the energy scale fixed by the
ultraviolet cutoff $\Lambda$,
one may be faced with non-Lorentz-covariant modifications of the
infrared energy spectrum that depend on the vacuum structure \cite{salehi}.
To indicate this type of infrared-ultraviolet connection, originating 
from vacuum condensation in effective quantum field theories, 
Volovik \cite{volo1} has introduced a very appropriate name: 
reentrant violations of special relativity in the low-energy corner. 
These are expected in a small shell of three-momenta, 
say $|{\bf{k}}| < \delta$, 
that only vanishes in the strict
local limit where $\Lambda \to \infty$ and an exact 
Lorentz-covariant energy spectrum
is re-obtained in the whole range of momenta. 

To understand the physical nature of these effects, let us assume that
the spontaneously broken phase of 
a $\lambda\Phi^4$ theory is a real Bose condensate
of physical spinless quanta. In the cutoff theory, these
are treated as `hard spheres' with a repulsive core $a\sim 1/\Lambda$
 analogously to the molecules of ordinary matter. Thus, 
in the limit of very long wavelengths ${\bf{k}} \to 0$, one
expects the lowest excitations to arise from small displacements of
the condensed quanta, that already `pre-exist' in the ground state,
giving rise to density fluctuations whose energy vanishes when
${\bf{k}} \to 0$.  

Notice the difference with the usual
empty vacuum state of a massive theory. There, 
a state with non vanishing three-momentum 
${\bf{k}} \neq 0$ can only be obtained
after the creation of one or more massive
quanta thus providing the physical argument for the existence of a 
mass-gap in the energy spectrum. 
On the other hand, 
in the case of a medium, the existence of density fluctuations
is a very general experimental fact that depends on the coherent response of
the elementary constituents to disturbances whose wavelength  
becomes larger than their mean free path. 
This leads to an universal description, the
`hydrodynamic regime', that does not depend on the details of the
underlying molecular dynamics and even on the nature of the elementary 
constituents. For instance, the same low-energy picture is expected in 
superfluid fermionic vacua \cite{volo2} that, as compared to the Higgs 
vacuum, bear the same relation of superfluid $^3$He to superfluid $^4$He.

Now, the basic macroscopic properties of Bose liquids
can be understood in terms of their long-wavelength excitations: phonons.
In fact, \cite{pita} 
"Any quantum liquid consisting of particles with integral 
spin (such as the liquid isotope $^4$He) must certainly have a spectrum of
this type...In a quantum Bose liquid, elementary excitations with small 
momenta ${\bf{k}}$  (wavelengths large compared with distances between atoms) 
correspond to ordinary hydrodynamic sound waves, i.e. are phonons. This 
means that the energy of such quasi-particles is a linear function of their
momentum". 

Therefore, 
quite independently of the Goldstone phenomenon, the energy spectrum 
of a Higgs condensate should terminate with 
an `acoustic' branch, say 
${E}({\bf{k}})=c_s|{\bf{k}}|$ for 
${\bf{k}} \to 0$, i.e. for momenta $|{\bf{k}}|< \delta$, where the 
associated wavelength 
${{2\pi}\over{\delta}}$ is 
larger than $ r_{\rm mfp}$, the mean free path for the condensed scalar 
quanta. 
\vskip 10 pt
{\bf 1.3} As anticipated, these gap-less modes, 
are quite unrelated to the usual Goldstone modes of spontaneously broken 
continuous symmetries and would
exist even in a Bose
condensate of strictly neutral bosons , such as the atoms
of $^4$He. In fact, in the quantum field
theoretical case, the translation from `field jargon to particle jargon', 
amounts to establish a well defined functional relation (see ref.\cite{mech}
and Sect.2)
$n=n(\phi^2)$ between the average particle density 
$n$ in the ${\bf{k}}=0$ mode and the average value of the scalar 
field $\phi$. Thus, Bose condensation is just a consequence of
the minimization condition of the 
 effective potential $V_{\rm eff}(\phi)$. This has absolute
minima at some values $\phi =\pm v \neq 0$ for which $n(v^2)=\bar{n}\neq 0$
\cite{mech}. 

In spite of its intuitive nature, this conclusion seems to be in
conflict with the standard analysis of the broken phase in a
one-component $\Phi^4$ theory. Indeed, this standard analysis predicts
a purely massive single-particle
energy spectrum $\sqrt{ {\bf{k}}^2 + M^2_H}$. Here
the parameter $M_H$, 
the Higgs boson mass, is defined from the quadratic shape of 
$V_{\rm eff}(\phi)$ at $\phi=\pm v$ and should be non-zero. 

However, as reviewed in Sect. 3, 
no real conflict with the intuitive picture of the broken phase as
a quantum liquid exists if one
takes into account the results of more formal analyses 
\cite{legendre,pmu} of the zero-4-momentum
inverse connected propagator
$G^{-1}(k=0)$ in the broken phase. 
Once the scalar condensate is not treated as
a purely classical c-number field, $G^{-1}(k=0)$
is a {\it two-valued} function and
includes the case $G^{-1}(k=0)=0$, as in a gap-less theory. 

More precisely, the existence of both a $G^{-1}_a(k=0)= M^2_H$ and a
$G^{-1}_b(k=0)= 0$ implies that there are
two possible types of excitations with the same quantum numbers but
different energies when the 3-momentum ${\bf{k}} \to 0$: 
a single-particle
massive one, with ${E}_a({\bf{k}}) \to M_H$, and a collective
gap-less one with 
${E}_b({\bf{k}}) \to 0$.  `A priori', they can both propagate 
(and interfere) in the broken-symmetry phase as in 
superfluid $^4$He, 
where the observed energy spectrum is due to the peculiar
transition from the `phonon branch' to the `roton branch' at a momentum scale 
$|{\bf{k}}_o|$ where
\BE
{E}_{\rm phonon}({\bf{k}}_o) \sim {E}_{\rm roton}({\bf{k}}_o)
\EE
This analogy supports the view that 
a scalar condensate is like
a real physical medium and can propagate different types of excitations. 
At the same time, 
deducing the detailed form of the energy spectrum 
that interpolates between 
${E}_a({\bf{k}})$ and ${E}_b({\bf{k}})$ is a formidable task. In fact, the
same problem in superfluid $^4$He, after
more than fifty years and despite the efforts of many theorists, notably
Landau and Feynman, has not been solved in a satisfactory way. 

Nevertheless, even without knowing the spectrum in full detail, one can draw
a certain number of conclusions. 
For instance, the gap-less, acoustic branch, say
${E}_b({\bf{k}})=c_s|{\bf{k}}|$, dominates for
${\bf{k}} \to 0$ and gives rise to a weak, 
attractive long-range force proportional to $1/c^2_s$
\cite{weak}. To this end, let us
consider the standard Fourier transform
\BE
D(r)=
\int {{d^3 {\bf{k}} }\over{(2\pi)^3 }}
 {{e^{ i {\bf{k}}\cdot {\bf{r}} } }\over{ E^2({\bf{k}}) }}
\EE
that, in QED, after replacing $E^2({\bf{k}})=|{\bf{k}}|^2$ as the appropriate
value for photon propagation, gives rise to 
the $1/r$ Coulomb potential. In our case, assuming
an energy spectrum with the following limiting behaviours :

~~~a) ${E}({\bf{k}}) \to {E}_a({\bf{k}}) = \sqrt{ {\bf{k}}^2 + M^2_H}$
        when $|{\bf{k}}|\to \infty $

~~~b) ${E}({\bf{k}}) \to {E}_b({\bf{k}}) = c_s |{\bf{k}}|$   
        when ${\bf{k}}\to 0 $

and using the Riemann-Lebesgue theorem on Fourier transforms \cite{goldberg}, 
one deduces the leading asymptotic behaviour for $r \to \infty$ 
\BE
\label{goldberg}
      D(r)= {{1}\over{4\pi c^2_s r}} (1+ {\cal O}(1/M_H r, 1/\delta r))
\EE
$\delta$ being the momentum scale associated with the transition between
the two regimes a) and b). 

To understand the value of $c_s$, we shall use in Sect.4
the formalism of ref.\cite{mech}. This leads to
a well defined hierarchy of scales
 $\delta \ll M_H \ll \Lambda$ that decouple in the continuum limit. 
Indeed, they are related through
\BE
           \Lambda \delta \sim M^2_H
\EE
so that the sound velocity is
\BE
           c_s \sim {{M_H}\over{\delta}} \sim {{\Lambda}\over{M_H}}
\EE
(in units of the light velocity $c=1$).

Therefore, the continuum limit can be defined in two equivalent ways. 
On one hand, one can require an exact Lorentz-covariant spectrum down to
${\bf{k}}=0$. This means
$r_{\rm mfp}\sim {{2\pi}\over{\delta}} \to \infty $ 
(in units of $\xi_H=1/M_H$)
so that the phonon wavelengths become larger than any finite scale.
When taken at face value, this corresponds to phonons of `infinite'
wavelengths that, with periodic boundary conditions, become
indistinguishable from the zero mode of the scalar field that defines the 
unperturbed condensate itself. 

On the other hand, the continuum theory can also be defined as
$\Lambda/M_H \sim c_s\to \infty$. In this other limit, 
the scalar condensate become inconpressible so that
the associated long-range 
force becomes instantaneous and of vanishingly small strength. Thus
the acoustic branch becomes unphysical since
$c_s |{\bf{k}}|$ exceeds the energy of the massive mode 
$\sqrt{ {\bf{k}}^2 + M^2_H}$ for any finite value of $|{\bf{k}}|$
(i.e. with the exception of the zero-measure set ${\bf{k}}=0$). 
\vskip 10 pt
{\bf 1.4} At the same time, in the cutoff theory, where 
the rigidity of the vacuum is very large but
not infinite, density fluctuations for $|{\bf{k}}| < \delta$ 
will propagate at a superluminal speed. 
Qualitatively, this same effect can be understood \cite{seminar,weak}
without looking at the energy 
spectrum but considering the relation between the pressure ${\cal P}$ and
the energy-density ${\cal E}$. In this case, treating the scalar condensate as
a zero-temperature perfect fluid with
${\cal P}= {-\cal E} + n d{\cal E}/dn$, one gets 
\BE
\label{cc1}
      {{c^2_s}\over{c^2}}= 
{{d {\cal P} }\over{ d {\cal E} }} =
{{n d^2 {\cal E} /dn^2  }\over{ d {\cal E}/dn }} 
\EE
Using the result $n=n(\phi^2) \sim \phi^2$ (see ref.\cite{mech} and Sect.2) 
and the `Maxwell construction' for the energy density, i.e.
${\cal E}(n)=V_{\rm eff}(\phi)$ for $\phi^2 > v^2$ and 
${\cal E}(n)=V_{\rm eff}(\pm v)$ for $\phi^2 \leq v^2$, the value of
$c^2_s/c^2$ can become arbitrarily large 
near the symmetry-breaking minima. 
For instance, approximating $V_{\rm eff}(\phi)$
with a standard double-well form 
$\sim (\phi^2-v^2)^2$, one finds 
\BE
\label{cc2}
         {{c^2_s}\over{c^2}}= 1 + {{v^2}\over{ \phi^2 -v^2}}
\EE
Although Eq.(\ref{cc1}) cannot be used {\it at} $\phi=\pm v$, where 
collisional effects from a finite mean free path are essential to obtain
a finite value of $c_s$ \cite{seminar}, the 
previous argument confirms that the condensed vacuum of 
a $\lambda\Phi^4$ theory
represents a very peculiar form of matter and can support nearly 
instantaneous density fluctuations. 

In this context, I observe that 
theoretical and experimental motivations \cite{recami,chiao}
show that the idea of waves with a superluminal group velocity
is physically meaningful. In particular, the possible existence
of superluminal density fluctuations in a condensed medium
poses no conceptual problem. For this reason, 
several authors \cite{poly,bludman1,keister} have considered
the possibility of media whose long-wavelength 
compressional modes have phase and group velocity 
${{{E}}\over{ |{\bf{k}}| }}=
{{d{E}}\over{ d|{\bf{k}}| }}=c_s>c$. In this sense, 
"..it is an open question whether 
${{c_s}\over{c}}$ remains less than unity when nonelectromagnetic forces are
taken into account" \cite{weisound}. 

In some case \cite{bludman1}, superluminal sound arises when
a large negative bare mass and a large positive self-energy 
combine to produce a small physical mass, just 
the situation expected for the quanta of 
the scalar condensate. More precisely, the physical
origin of the superluminal sound is traced back to the 
asymmetric role of mass renormalization: it subtracts out self-interaction
energy without altering the tree-level
interparticle interactions that contribute
to the pressure. Therefore, for (nearly) point-like particles in three spatial
dimensions and sufficiently small values of their mass, 
matter can become superluminal, i.e. $d{\cal P}/d{\cal E} > 1$. 

\vskip 10 pt
{\bf 1.5} As the condensate compressional modes give rise to  very 
weak interactions suppressed by a $1/c^2_s$ factor, 
and in connection with the
 hierarchical pattern of scales $\delta \ll M_H \ll \Lambda$,  
I shall try to combine these ingredients 
to connect the Fermi and Planck scales, thus providing a possible solution
of the `hierarchy problem'. 

To this end, the first
minimal requirement will consist in deriving in Sect.5
Newton's theory 
of gravity from the underlying physics of the Higgs condensate.
In fact, up to now, 
"How relevant the Planck energy is to elementary particle physics has not
really been established. It's merely a number with the dimension of a mass 
that comes out of Newton's theory of gravity. Call it Planck mass if you
wish. It may or may not play a fundamental role" \cite{glashow}.

In addition, as a second basic requirement, the phonon dynamics should be 
`geometrizable', i.e. re-absorbable into an effective geometry
that agrees, at least for weak field, 
with the known metric structure of General Relativity. This analysis will
be presented in Sect.6.

Finally, Sect.7 will
contain a summary of the paper and a discussion of other possible
consequences of my approach.

\section{The Bose-condensate picture of symmetry breaking}

I shall now start by
resuming the main results of ref.\cite{mech} in the case of
a one-component $\lambda \Phi^4$ theory, a system 
 where the condensing quanta are just neutral spinless 
particles, the `phions'. 

One starts by quantizing 
the scalar field $\Phi({\bf{x}})$ in terms of 
$a_{\bf k}$, $a^{\dagger}_{\bf k}$, the 
annihilation and creation operators for the elementary phions
whose `empty' vacuum state $|o\rangle$ is defined through
$a_{\bf k}|o\rangle=\langle o|a^{\dagger}_{\bf k}$=0. 

The phion system is assumed to be contained within a finite box of volume $\vol$ 
with periodic boundary conditions.  There is then a discrete set of allowed 
modes ${\bf k}$.  In the end one takes the infinite-volume limit and the 
summation over allowed modes goes over to an integration: 
$\sum_{\bf k} \to \vol \int d^3k/(2 \pi)^3$.  In this way, the scalar field
is expressed as
\BE
\label{phime}
\Phi({\bf x}) = 
\sum_{\bf k} \frac{1}{\sqrt{2 \vol E_k}} 
\left[ a_{\bf k} {\rm e}^{i {\bf k}.{\bf x}} + 
a^{\dagger}_{\bf k} {\rm e}^{-i {\bf k}.{\bf x}} 
\right] ,
\EE
where $E_k=\sqrt{{\bf k}^2 + m^2_\Phi}$, $m_\Phi$ 
being the physical, renormalized phion mass.

Bose condensation means that in the ground state
 there is an average number $\num_0$ of phions in the ${\bf k}=0$ 
mode, where $\num_0$ is a finite fraction of the 
total average number $\num$
\BE
         \num=\langle 
\sum_{\bf{k}} a^{\dagger}_{ {\bf{k}} } a_{ {\bf{k}} } \rangle
\EE
At zero temperature, if the gas is dilute, almost all the particles are in 
the condensate; $\num_0{\small (T=0)} \sim \num$.  In fact, the fraction 
which is not in the condensate (`depletion')
\BE
\label{deple}
            D= 1- {{\num_0}\over{\num}} ={\cal O}(\sqrt{n a^3})\ll 1
\EE
is a phase-space effect
that becomes negligible for a very dilute system. 
In Eq.(\ref{deple}) we have introduced the phion density
\BE
\label{density}
                        n= {{\num}\over{\vol}}
\EE
and the phion-phion scattering length 
\BE
\label{aa}
a = \frac{\lambda}{8 \pi m_\Phi}  
\EE
defined, in the limit of zero-momentum scattering, from the dimensionless
scalar self-coupling $\lambda$ and the phion mass. 

Therefore, for a very dilute system, where, to a first approximation, one 
neglects the residual operator part of $a_{ {\bf{k}}=0}$, one gets 
$a^{\dagger}_{ {\bf{k}}=0} a_{ {\bf{k}}=0} \sim \num$ and so, 
 $a_{ {\bf{k}}=0}$ can be identified with
the c-number $\sqrt{\num}$ (up to a phase factor).  In this way
\BE
\label{phn}
\phi = \langle \Phi \rangle = \frac{1}{\sqrt{2 \vol m_\Phi}} 
\langle (a^{\dagger}_{ {\bf{k}}=0}+ a_{ {\bf{k}}=0}) \rangle \sim  
\sqrt{\frac{2\num}{\vol m_\Phi}}.
\EE
or 
\BE
\label{neq}
n(\phi^2) \sim \half m_\Phi \phi^2,
\EE
With this identification, setting 
$a^{\dagger}_{ {\bf{k}}=0}= a_{{\bf{k}}=0}=\sqrt{\num}$ is equivalent to 
shifting the quantum field $\Phi$ by a constant term $\phi$. 

Finally, using Eq.(\ref{neq}), 
the energy density ${\cal E}={\cal E}(n)$
 can be translated into the
effective potential 
\BE
V_{\rm eff}(\phi)={\cal E}(n)
\EE
Therefore, provided
the effective potential of the (cutoff) $\lambda\Phi^4$ theory 
exhibits non-trivial absolute minima $\phi=\pm v \neq 0$
for real and non-negative values of 
$m_\Phi$, spontaneous symmetry breaking is equivalent to Bose condensation 
with an average density $\bar{n}=n(v^2)$.
In other words, Bose condensation requires spontaneous symmetry breaking in
(cutoff) $\lambda\Phi^4$ theories to be a 
first-order phase transition, differently
from the second-order picture of
standard renormalization-group-improved 
perturbation theory.

The problematic aspects associated with standard perturbation theory 
in `trivial' $\lambda\Phi^4$ theories \cite{book}
have been discussed in \cite{trivpert}
and the whole issue has been reviewed in detail in ref.\cite{mech}.
The result of those analyses, as well as of
refs.\cite{zeit,plb}, is the following. 
 There is a class of `triviality compatible'
approximations to $V_{\rm eff}$ (one-loop potential, gaussian and 
post-gaussian \cite{ritschel} calculations of the
Cornwall-Jackiw-Tomboulis \cite{cjt} effective potential for composite
operators), i.e. where the fluctuation field is described by an effective 
quadratic hamiltonian, in which spontaneous symmetry breaking 
is described as an 
infinitesimally weak first-order phase transition. This means that, 
in this class of 
approximations, differently from the standard perturbative predictions, 
the phase
transition is first-order in the cutoff theory and becomes asymptotically
second-order when approaching the continuum limit \cite{notecw}.

The physical ingredient to understand this result
consists in the observation \cite{mech} that the 
phion-phion interaction is not always repulsive. Besides the tree-level
contact 
$+\lambda\dl$ potential, there is an induced attraction 
$-\lambda^2 {{e^{-2m_\Phi r}}\over{r^3}}$ from the {\it ultraviolet-finite}
 parts of the one-loop
graphs (see also ref.\cite{ferrer}) that, differently from the 
standard ultraviolet divergences, cannot be reabsorbed into a perturbative
redefinition of the tree-level coupling. Actually, 
the two effects replicate themselves to all orders \cite{mech}
since, for each ultraviolet divergent
higher-order graph that redefines the strength of the tree-level potential,
say $\lambda \to \lambda_{\rm eff}$, 
there is a corresponding effect that redefines in the same way, 
$\lambda^2\to \lambda^2_{\rm eff}$, the strength of the attractive tail.

Thus, the long-range attraction persists to all orders
and is responsible for the failure of 
ordinary renormalization-group-improved perturbation theory when 
approaching the phase transition. In fact,  
when making $m_\Phi$ smaller and smaller, the
long-range attractive tail extends over regions that become larger 
and larger in units of the length scale associated with the repulsive core
$+\lambda_{\rm eff}\dl$. As a consequence, 
the corresponding graphs, when taken into account consistently in the 
effective potential, can compensate for the effects of
both the short range repulsion and of
the non-zero phion mass. In this 
situation, the perturbative empty vacuum state $|o\rangle$, although
locally stable, is not globally stable
and, in a regime where
 ${{m^2_\Phi}\over{na}}= {\cal O}(\lambda)$,
the lowest energy state becomes a state with a non-zero density of phions that
are Bose condensed in the zero-momentum state. 

I emphasize that this weakly first-order scenario of symmetry breaking 
is discovered in a {\it class} of approximations to the effective 
potential, just those that are consistent with the assumed exact `triviality'
properties of the theory in 3+1 dimensions. 
 In any case, it can be objectively tested against
the standard perturbative predictions
based on a second-order phase transition. In fact, one can run 
numerical simulations near
the phase transition region and compare the predictions of 
refs.\cite{zeit,plb} with the conventional existing two-loop or 
renormalization-group-improved forms of the effective potential. 
As the existing lattice data  \cite{agodi,fiore} 
support unambiguosly the alternative picture of refs.\cite{zeit,plb,mech},
I shall take this result as an additional motivation 
for pursuing further the Bose-condensate picture of symmetry breaking.

\section{The excitations of the scalar condensate}

~~~{\bf 3.1}~ As anticipated in the Introduction, and on the base of 
refs.\cite{legendre,pmu}, the
excitation spectrum of the broken phase has a two-branch structure due
to the existence of two different solutions, 
$G^{-1}_a(k=0)=M^2_H$ and
$G^{-1}_b(k=0)=0$, 
for the connected zero-4-momentum
inverse propagator 
$G^{-1}(k=0)$. In this way, even without solving the integral equation 
for $G(k)$, one can deduce that
for ${\bf{k}} \to 0$, 
there are two excitations of the scalar condensate: a massive one with 
energy 
$E_a({\bf{k}}) \to M_H$ and a gap-less one with energy
$E_b({\bf{k}}) \to 0$. They will be analyzed in the following.
\vskip 10 pt
{\bf 3.2}~ The parameter $M_H$, entering the massive solution, 
is obtained after introducing the absolute minima $\phi=\pm v$ 
of the effective potential
$V_{\rm eff}(\phi)$.
As such, $M_H$ is extracted
from the propagator as defined from the Dyson sum
of one-particle {\it irreducible} graphs only.
In this way, using simple tree-level relations, one 
can relate $M_H$ to $n$ and $a$ through
 ($ \bar {n}= n(v^2)$)
\BE
\label{mh}
                     M^2_H \sim \lambda v^2 \sim \bar{n}a 
\EE
In this context, one should notice
a peculiar point concerning the relation between $M_H$ and the physical 
vacuum field, say $v^2_R$, that in the Standard Model is then related to the
Fermi constant ($G_F\sim 1/v^2_R$).
Up to now, the normalization of $\phi$ has been determined from the
vacuum expectation value of the 
bare scalar field $\Phi(x)$ defined at the cutoff scale. 
On the other hand, when using the standard condition
\BE
\label{phys}
\left.{{d^2V_{\rm eff}(\phi_R)}\over{d\phi^2_R}}\right|_{\phi_R=\pm v_R}
=M^2_H
\EE
to define the physical normalization 
of the vacuum field, say 
\BE
\label{zz}
v_R= {{v}\over{ \sqrt{Z_\phi} }}
\EE
one has to account for the non-trivial renormalization effect discussed
in refs.\cite{zeit,plb,mech} and observed in lattice computations
\cite{cea1,cea2} of the zero-momentum susceptibility 
$\chi_{\rm latt}$ in the broken phase. 
This effect is a direct consequence of Bose condensation, that gives a special
role to the vacuum condensate. In fact, 
differently from the rescaling
of the fluctuation field, as defined from the K\"allen-Lehmann decomposition, 
 say
$Z_{\rm prop} = 1 - |{\cal O}(\lambda)|$, the vacuum field re-scales 
non-perturbatively as
\BE
\label{zphi}
              Z_\phi\sim {{1}\over{\lambda}}
\EE
Therefore, one always ends up with the cutoff-independent relation
\BE
\label{mh2}
                     M^2_H \sim v^2_R
\EE
even in the limit of
a vanishingly small coupling $\lambda$. For this reason, the 
existence of $Z_\phi$ represents the basic ingredient to reconcile
spontaneous symmetry breaking with the generally accepted `triviality' of
$\lambda\Phi^4$ theory in 4 space-time dimensions \cite{book}. In this way, 
the continuum limit $\lambda \to 0$ 
corresponds to free-field fluctuations with $Z_{\rm prop}=1$ but
finite values of $M_H$ and $v_R$. Physically, 
these two quantities can be thought as arising from equivalent 
 phion condensates with smaller and smaller scattering lengths $a$ but
increasingly large particle density $\bar{n}$ 
such that $\bar{n}a$ remains constant. 
\vskip 10 pt
{\bf 3.3}~ Let us now consider the gap-less solution $G^{-1}_b(k=0)=0$. 
Following the discussion presented in the Introduction, it
would be naturally understood as the
acoustic branch, i.e.  $E_b({\bf{k}})= c_s|{\bf{k}}|$, associated
with the long-wavelength oscillations of the scalar condensate. 
This dominates in the infrared region and
represents the physical mechanism for a new long-range force \cite{weak}. 

I observe that the gap-less solution
is found when treating the 
zero-mode of the scalar field as a genuine quantum degree of freedom \cite
{legendre} over which to perform the last functional integration
to compute the zero-four-momentum propagator. 
 However, it can also be discovered
diagrammatically \cite{pmu} provided one
includes in the analysis of the propagator
a new class of graphs: the one-particle
{\it reducible} zero-momentum tadpole graphs. These 
originate from the vacuum source 
$J(\phi)={{dV_{\rm eff}(\phi)}\over{d\phi}}$ and
are connected to the other parts
of the diagrams through zero-momentum propagators.  As such, this class of
graphs can be considered
a manifestation of the quantum nature of the scalar condensate 
and, traditionally, 
is not included in the standard perturbative expansion {\it at} $\phi=\pm v$
where $J(\pm v)=0$.

However, in an all-order analysis for arbitrary $\phi$,  where the limit
$\phi \to \pm v$ is only taken at the end of the calculation, 
the one-particle reducible 
tadpole graphs have to be included to obtain the correct propagator.
 In an intuitive mechanical analogy, 
$J(\phi)$ represents an infinitesimal driving force. Thus it will not produce
any observable effect, unless the mass of a body vanishes 
when $J(\phi) \to 0$. In our case, the tadpole graphs
are attached to the other parts of the
diagrams through zero-momentum propagators so that
the `mass' of our body is just the
inverse zero-momentum propagator. 
 In this sense, 
ignoring such contributions (or including their effect in a pure
perturbative way as $G(k=0)={{1}\over{M^2_H}}+...$) one neglects
the potentially singular nature of 
$G(k=0)$ 
and only finds the massive solution. 

Addressing to ref.\cite{pmu} for the details, 
the tacit assumption at the base
of the standard approach is better 
illustrated with a very simple example. Consider the quadratic equation 
\BE
           f^{-1}(x)= 1 +x^2 -g^2 x^2 f(x)
\EE
for $g^2 \ll 1$. The analogy with the problem of the one-particle reducible
zero-momentum tadpole graphs is established when comparing
$f(x)$ with $G(k=0)$ at a given 
value of $\phi$ and comparing the limit $x \to 0$ 
with the limit $J(\phi) \to 0$. Standard
perturbation theory is based on the iterative structure 
$f_{\rm reg}=1/(1+x^2) + {\cal O}(g^2)$ 
that provides a class of 
solutions that
are regular for $x \to 0$ where $f_{\rm reg}(0)=1$. This corresponds to the
massive propagator as defined from the one-particle irreducible graphs only.
 On the other hand, the singular solution 
$f_{\rm sing}\sim 1/g^2 x^2$ corresponds to 
a divergent zero-momentum propagator when $\phi \to \pm v$. 
This can only be discovered by retaining
the full non-linearity of the problem where the zero-momentum 
propagators joining to the vacuum sources are not approximated perturbatively. 

Since the existence of two solutions for $G(k=0)$
depends on how one treats the 
general case $\phi \neq \pm v$, and since we know that one can translate
from $\phi^2$ to $n$, it may be interesting
 to establish an analogy with the possible ways of defining the 
theory at an arbitrary particle density $n\neq \bar{n}$. In this 
way, one can obtain
an explicit form for a two-branch energy spectrum
without any need to include the second-quantized 
counterpart of the one-particle reducible zero-momentum tadpole graphs. 

To this end, I observe that for $n \neq \bar{n}$, 
the phion condensate cannot sustain itself. In fact, strictly speaking
$\mu_c=d {\cal E}/dn =0$ only for
$n = \bar{n}$. Thus, to consider the general case $n \neq \bar{n}$, 
one may introduce a fictitious
 external chemical potential $\mu_c$ that will be sent to
zero at the end of the calculation. 
 In this case, still within the standard 
Bogolubov approximation, 
 one can use Eq.(4.17) of ref.\cite{mech} and obtain the
energy spectrum 
\BE
\label{spectrum}
   E({\bf{k}})= ( \sqrt{   {\bf{k}}^2 + m^2_\Phi } - \mu_c) \sqrt{ 1 + 
{ {  8\pi n a }\over{  \sqrt{  {\bf{k}}^2 + m^2_\Phi} 
(\sqrt{ {\bf{k}}^2 + m^2_\Phi } - \mu_c) }} }
\EE
Now, in the semi-classical approximation of ref.\cite{kapusta} 
for a temperature $T=0$, one would predict $ m_\Phi \leq \mu_c$ 
as the condensation condition with a positive physical mass of the
scalar quanta (compare with Fig.3 of ref.\cite{kapusta}). Therefore, 
by requiring a physical non-negative $E({\bf{k}}=0)$, 
i.e. $\mu_c=m_\Phi$, and taking the limit $m_\Phi \to 0$ one obtains
\BE
   E({\bf{k}}) = \sqrt{ {\bf{k}}^2+ 8\pi na }
\EE
On the other hand, by first taking the limit ${\bf{k}} \to 0$, 
for any arbitrarily small but non-zero
$m_\Phi$, one gets
\BE
\label{ex1}
 \lim_{ {\bf{k}}\to 0}  
E({\bf{k}})\sim { {|{\bf{k}}|}\over{2m_\Phi}} \sqrt {16 \pi na}
\EE
while still 
\BE
\label{ex2}
   E({\bf{k}}) \sim \sqrt{ {\bf{k}}^2+ 8\pi na }
\EE
at larger $|{\bf{k}}|$. In this case the infrared limit is
different and one finds, even for $n=\bar n$,  an explicit realization 
of a two-branch energy spectrum. 

As in the case of the zero-momentum 
tadpole graphs, there is a subtlety associated with two non-commuting limits. 
In the latter case, the limit $m_\Phi \to 0$ yields a massive spectrum
with mass $M^2_H \sim \bar{n}a$ with
the exception of the zero-measure set ${\bf{k}}=0$ or, more precisely, with the 
exception of a region $|{\bf{k}}| < \delta$ where 
\BE
\label{mphidelta}
\delta\sim m_\Phi
\EE
is infinitesimal in units of $M_H= \sqrt{8 \pi n a}$. 

As mentioned in Sect.2, an infinitesimal value of the ratio 
$m_\Phi/\sqrt{na}$ is found in the weakly first-order scenario
of refs.\cite{zeit,plb,mech}. Actually, as it will be
shown in Sect.4, relation
(\ref{mphidelta}) is precisely the one expected when $\delta$ is taken
to represent
the inverse mean free path for the condensed phions. Therefore, the  
infrared trend Eq.(\ref{ex1}) provides an explicit realization of the
`acoustic regime' mentioned in the Introduction. 

\vskip 10 pt
{\bf 3.4}~ I conclude this section with some remarks concerning the eigenmode
expansion of the scalar field in the broken phase. Beyond perturbation theory, 
due to the different physical status of the vacuum field
and of its quantum fluctuations, 
the field re-scaling cannot be given as an overall `operatorial'
condition \cite{agodi} (of the type, say, $\Phi_R(x)=\sqrt{Z} \Phi(x)$).
In fact, Eqs.(\ref{phys}) and (\ref{zphi}), used to define
the physical normalization of the vacuum field, coexist with a free-field
normalization for the massive branch of the
fluctuating field (i.e a value $Z_{\rm prop}=1$). Actually, 
Eq.(\ref{phys}) represents the standard condition for 
a smooth zero-4-momentum limit of a free-field propagator $1/(k^2 + M^2_H)$
with mass $M_H$. Ignoring for a moment the existence
of the gap-less branch, such physical representation of the scalar field
amounts to write (phys=`physical') 
\BE
\label{conve1}
\Phi_{\rm phys}(x)= v_R + H(x)
\EE 
where
\BE
\label{conve2}
{H}(x)=
\sum_{ {\bf {k}} }
\frac{1} { \sqrt{2 \vol {E}_k } } 
\left[  \tilde{H}_{\bf k}    {\rm e}^ { i ({\bf k}.{\bf x} -{E}_k t) } + 
(\tilde{H}_{\bf k})^{\dagger} {\rm e}^{-i ({\bf k}.{\bf x} -{E}_k t)} 
\right]
\EE
with ${E}_k=\sqrt{{\bf{k}}^2 + M^2_H}$.

In the usual approach, where the Higgs vacuum is treated as a purely classical
c-number field and the Higgs boson is represented 
as a purely massive quantum field, 
the decomposition Eq.(\ref{conve2}) is adopted even when dealing 
with wavelengths $2\pi/|{\bf{k}}|$ (say one meter)
that are infinitely larger than $\xi_H=1/M_H$. 
In our case, however, to account
for the existence of two types of fluctuations, it is convenient to
introduce two fields, $h(x)$ and $H(x)$, 
whose Fourier components are separated by
the end of the acoustic branch, say $|{\bf{k}}|= \delta$. 
In this simplified approach, where 
the energy spectrum is approximated by
two independent branches and no attempt is made to
find an appropriate 
form of matching condition, Eqs.(\ref{conve1}) and (\ref{conve2}) are 
replaced by the alternative expression
\BE
\label{phime2}
\Phi_{\rm phys}(x) = v_R + {h}(x) + {H}(x)
\EE 
with 
\BE
\label{hh}
{h}(x)=
\sum_ { | {\bf {k}}| < \delta }  
\frac{1} { \sqrt{2 \vol {E}_k } } 
\left[  \tilde{h}_{\bf k}    {\rm e}^ { i ({\bf k}.{\bf x} -{E}_k t) } + 
(\tilde{h}_{\bf k})^{\dagger} {\rm e}^{-i ({\bf k}.{\bf x} -{E}_k t)} 
\right]
\EE
and
\BE
\label{H2}
{H}(x)=
\sum_{ |{\bf {k}}| > \delta }  
\frac{1} { \sqrt{2 \vol {E}_k } } 
\left[  \tilde{H}_{\bf k}    {\rm e}^ { i ({\bf k}.{\bf x} -{E}_k t) } + 
(\tilde{H}_{\bf k})^{\dagger} {\rm e}^{-i ({\bf k}.{\bf x} -{E}_k t)} 
\right]
\EE
where ${E}_k=c_s|{\bf{k}}|$ for $|{\bf{k}}| < \delta$, 
${E}_k=\sqrt{{\bf{k}}^2 + M^2_H}$ for $|{\bf{k}}| > \delta$, 
with $c_s \delta \sim M_H$. The more conventional 
Eqs.(\ref{conve1}) and (\ref{conve2})
are reobtained in the limit $\delta \to 0$ (or $c_s \to \infty$ for a fixed
$M_H$) where $h(x)$ disappears and the broken phase has
only massive excitations.

\section{A hierarchy of scales in the condensate}

Let us now consider the pattern of scales associated with the 
phion condensate. In condensed matter, 
the transition between acoustic branch and single-particle
spectrum corresponds to their matching at 
a momentum scale set by the inverse 
mean free path for the elementary constituents. As anticipated, 
in the scalar condensate, this defines
a momentum $\delta$ where  $E_a(\delta) \sim E_b(\delta)$ or
\BE
\label{match}
               c_s \delta \sim \sqrt {\delta^2 + M^2_H}
\EE
with
$ {\delta} \sim {{1}\over{ r_{\rm mfp} }}$, $r_{\rm mfp}$ being
 the phion mean free path \cite{kine,seminar}
\BE
\label{mfp}
r_{\rm mfp} \sim {{1}\over{ \bar{n}a^2}}
\EE
Now,  assuming `triviality' 
the scattering length $a$ should vanish in the 
continuum limit of the theory. In this limit, the associated momentum scale
\BE
\Lambda \equiv 1/a
\EE
diverges in units of $M_H$ 
and, therefore, can be used to define a far ultraviolet scale
of the theory up to which phions can be treated as `hard spheres'. This
yields
\BE
\label{tt}
   t= {{\Lambda}\over{M_H}} \sim \sqrt{  {{1}\over{ \bar{n} a^3}} } 
\EE
and 
\BE
\label{epsilon}
{{1}\over{c_s}}\sim {{\delta}\over{M_H}} \sim 
\sqrt { \bar{n}a^3} 
\EE
Therefore, when $t \to \infty$ one gets an infinitely dilute
Higgs condensate where $\bar{n}a^3 \to 0$ and 
a hierarchy of scales $\delta \ll M_H \ll  \Lambda$ with
\BE
\label{golden}
                 \Lambda \delta \sim M^2_H
\EE
As anticipated in the Introduction, 
the order of magnitude of the
`sound' velocity
\BE
\label{cs}
       c_s \sim {{M_H}\over{\delta}} \sim {{\Lambda}\over{M_H}}
\EE
is much larger than unity (in units of $c=1$). 
Actually, $c_s$ {\it must} diverge
in the continuum limit where the vacuum has an infinite rigidity and the 
energy spectrum becomes
Lorentz covariant down to ${\bf{k}}=0$. In fact, formally, 
${\cal O}({{\delta}\over{M_H}})$ vacuum-dependent corrections are
equivalent to 
 ${\cal O}({{M_H}\over{\Lambda}})$ effects and these have always been
neglected when discussing \cite{nielsen} how
Lorentz covariance emerges in effective theories when removing
the ultraviolet cutoff. 

Finally, I observe that the order of magnitude relation 
Eq.(\ref{mphidelta}) is found naturally when the momentum scale $\delta$ is
identified with the inverse mean free path for the condensed phions. In fact, 
from refs.\cite{zeit,plb,mech}, one deduces that spontaneous symmetry
breaking sets in for infinitesimally small values of the ratio $m_\Phi/M_H$, 
precisely 
\BE
                  {{m^2_\Phi}\over{ \bar{n}a }} \sim \lambda \sim m_\Phi a
\EE
Therefore, using Eq.(\ref{mfp}), one finds 
\BE
\label{mphirmfp}
       m_\Phi \sim \bar{n} a^2 \sim {{1}\over{ r_{\rm mfp} }}
\EE
as anticipated.

Summarizing the previous results, we find that
in the local limit of the theory, where $\Lambda/M_H \to \infty$, 
the energy spectrum $E({\bf{k}})$ reduces to 
$\sqrt{ {\bf{k}}^2 + M^2_H}$ 
in the whole range of $|{\bf{k}}|$ (with the exception of
the zero-measure set ${\bf{k}}=0$). In the cutoff theory, however, 
one should expect infinitesimal deviations
in an infinitesimal region of three-momenta. Therefore, it is natural
to ask what the word {\it infinitesimal} might actually 
mean in the physical world. For instance, assuming
$\Lambda=10^{19}$ GeV and $M_H=250$ GeV, a scale 
$\delta\sim {{M^2_H}\over{\Lambda}}\sim 10^{-5}$ eV, for which
${{\delta}\over{M_H}}\sim 4\cdot10^{-17} $, 
 might well represent the physical 
realization of a formally infinitesimal quantity. If this were
the right order of magnitude, 
the collective density fluctuations of the Higgs vacuum
have wavelengths $> {{2\pi}\over{\delta}}$ 
thus ranging from about a centimeter up to infinity.
 
At the same time, the associated long-range force Eq.(\ref{goldberg}) would
correspond to a nearly instantaneous interaction 
(since $c_s\sim {{M_H}\over{\delta}}= {\cal O}(10^{16})$ 
in units of the light velocity) with a typical
strength ${{1}\over{c^2_s}} \sim {{M^2_H}\over{\Lambda^2}} =
{\cal O}(10^{-33})$. However small, 
this strength is non-vanishing and these interactions can 
 play a physical role over distances that are infinitely 
larger than any elementary particle scale.

\section{A weak attractive long-range force}

I shall now exploit
the idea that $\Lambda=1/a$ can be used to mark 
the Planck scale, $M_H \sim v_R$ can be taken to indicate
the Fermi scale $G_F\equiv 1/v^2_R$ so that $\delta \sim 10^{-5}$ eV.
As anticipated in the Introduction, for the validity of this identification, 
the long-range interactions associated with the longitudinal waves Eq.(\ref{hh})
should represent the physical mechanism at the base of Newtonian gravity.  
The possibility that, as it happens in rotating liquids \cite{chandra} or
in superfluid $^4$He \cite{wilks,tilley}, 
there might be other kind of waves (as transverse waves that cannot
be described in terms of a single scalar function) 
goes beyond the scope of this
paper and will only be briefly addressed as a concluding remark at the end
of Sect.6.

Starting from the simple representation based 
on Eq.(\ref{phime2})-(\ref{H2}), 
with $ \delta \sim 1/r_{\rm mfp}$, one should find
an effective macroscopic
interaction to describe the coupling of $h(x)$. 
By {\it macroscopic}, I mean that, starting, as in the Standard Model, from a 
model field theory for $\Phi_{\rm phys}$, one has to 
fill a gap of many orders of magnitude and find the global
coupling of $h(x)$ to matter over scales of linear 
size $r_{\rm mfp}$ 
or larger. As $r_{\rm mfp}= {\cal O}(1)$ centimeters, 
these lengths correspond to classical physics. Therefore, 
one has to take a suitable limit
where the relevant degrees of freedom are expressed in terms of
classical motions. To derive such a global coupling, one can argue as follows.

Let me first observe that in an exact Lorentz-invariant theory
the simplest possible coupling between ordinary matter 
and a (dimensionless) scalar field $\varphi(x)$ 
is through $T^\mu_\mu(x)$, 
the trace of the energy momentum tensor. In this case, 
up to terms in the derivatives of $\varphi$ and up to
higher powers in the $\varphi$-field, one gets 
the simple interaction lagrangian 
\BE
\label{trace}
              {\cal L}_{\rm int}= -T^\mu_\mu(x) \varphi(x)
\EE
with
\BE
\label{tmumu}
         T^{\mu}_{\mu}(x) \equiv \sum_n 
{ { E^2_n -   {\bf{p}}_n  \cdot  {\bf{p}}_n  } \over{E_n}} 
\delta^3 ( {\bf{x}} - {\bf{x}}_n(t) ) 
\EE
The normalization in Eq.(\ref{trace}) 
 corresponds to the standard form of the action 
\BE
\label{cla}
               S_{\rm int}= \int d^4 x {\cal L}_{\rm int}= 
- \sum_n M_n \int ds_n \varphi(x)
\EE
for point-like particles interacting with a scalar field $\varphi(x)$ where
$x_\mu=x_\mu(s_n)$ and
$ds_n=dt_n \sqrt{1 - {\bf{v}}^2_n}$ denotes the proper-time element of
the n-th particle.

Now, I shall assume that the field
$\varphi(x)$ represents the long-wavelength component of the full 
scalar field $\Phi_{\rm phys}(x)$ in Eqs.(\ref{phime2}) and (\ref{hh}), i.e.
\BE
\varphi(x)\equiv {{h}\over{v_R}}
\EE
so that its Fourier decomposition will only contain wavelengths that are 
larger than one centimeter or so. I shall also assume that
$\varphi(x)$ is the only non-Lorentz-invariant effect in the theory and
that the fundamental
couplings of $\Phi_{\rm phys}(x)$ are formally
Lorentz invariant. 

As a definite physical framework to obtain Eq.(\ref{trace}), 
I shall follow ref.\cite{weak}, by considering
a microscopic scalar-fermion yukawa coupling 
\BE
\label{yu}
{\cal L}_{\rm yukawa}
=-g_f \bar {\psi}_f \psi_f \Phi_{\rm phys}
\EE
In this case, 
using Eq.(\ref{phime2}) (and dropping the $H$-part), one obtains
\BE
{\cal L}_{\rm yukawa}=
-m_f \bar{\psi}_f\psi_f 
(1 +\varphi)
\EE
where 
\BE
\label{mf}
m_f=g_f v_R
\EE
Finally, if the field
$\psi_f$ describes a sharply localized wave packet
with momentum ${\bf{p}}$, i.e. such that
$\int d^3 x \langle \bar{\psi}_f\psi_f\rangle= m_f/ \sqrt{ {\bf{p}}^2 + m^2_f }$ and such
that $\varphi$ does not vary appreciably over the
localization region, the interaction in Eq.(\ref{yu})
produces the classical action in
Eq.(\ref{cla}) for a particle of mass $m_f$. 

In this particular case, where the elementary fermions have no 
interaction with other fields, the macroscopic global coupling 
of $\varphi(x)$ will be described as a formally 
Lorentz-invariant coupling of the type in 
Eqs.(\ref{trace}) and (\ref{cla})
and the mass parameters in Eq.(\ref{cla}) coincide 
with the mass in Eq.(\ref{mf}). 

An analogous result follows
when the fermions in Eq.(\ref{yu}) have
additional couplings to other fields.
Provided these other interactions are
Lorentz-invariant, $\varphi(x)$ will always couple to a classical,  
Lorentz-invariant, dimension-four variable depending on all positions and
velocities needed to describe the classical system. In this way,
 again, the basic coupling is through $T^\mu_\mu$. However, 
in this more general situation, 
there is no reason for the 
mass of the classical particles 
in Eq.(\ref{cla}) to coincide with any of
the mass parameters $m_f$ of the 
microscopic yukawa interactions in Eq.(\ref{mf}). 

To better understand this result, 
I'll concentrate on quarks, whose relevant additional couplings
are due to their QCD interactions, 
starting from the unperturbed situation where $\varphi=0$. In this case,
one can assume that
the mutual interactions among the various condensates give
rise to suitable relations arising from the minimization of the overall 
energy density. One can express these relations as
\BE
        \alpha_s\langle F^a_{\mu\nu}F^a_{\mu \nu}\rangle_o= c_1 v^4_R
\EE
and 
\BE
            m^{(o)}_q \langle \bar{\psi}_q \psi_q \rangle_o= c_2 v^4_R
\EE             
where $c_1$ and $c_2$ are dimensionless numbers. 
 
Now, let us consider
an external perturbation that induces long-wavelength oscillations of 
the scalar condensate so that
$v_R \to v_R(1+ \varphi)$. Let us also assume that QCD has only 
short-range fluctuations whose wavelengths are infinitely 
smaller than $r_{\rm mfp} \sim {\cal O}(1)$ centimeters. In this situation, 
$v_R(1+ \varphi)$ can be considered a new `local vacuum' that, however, 
extends over an infinitely large region on the QCD scale and
to which the quark masses and the relevant
expectation values of the gluon and quark operators have to be
adjusted. This amounts to write
\BE
        \alpha_s\langle F^a_{\mu\nu}F^a_{\mu \nu}\rangle= c_1 
v^4_R(1+\varphi)^4
\EE
and 
\BE
            m_q \langle \bar{\psi}_q \psi_q \rangle= 
c_2 v^4_R(1+\varphi)^4
\EE             
The overall result is equivalent to a rescaling of the QCD scale 
parameter 
\BE
\Lambda_{\rm QCD} \to \Lambda_{\rm QCD}(1+ \varphi)
\EE 
of the chiral condensate 
\BE
\langle \bar {\psi_q} \psi_q \rangle \to 
\langle \bar {\psi_q} \psi_q \rangle (1+ \varphi)^3
\EE
and of the quark masses
\BE
               m_q \to m_q (1+ \varphi)
\EE
so that the nucleon mass $m_N$, 
arising from the expectation
value of both quarks and gluon condensates, undergoes the same rescaling
\BE
               m_N \to m_N (1+ \varphi)
\EE
In this way, regardless of the detailed relation of the 
classical $T^\mu_\mu$ Eq.(\ref{tmumu}) to the fundamental fields, 
the only remnant of the non-Lorentz-invariant nature
of the vacuum is represented
 by the {\it free} lagrangian
of the $\varphi$-field entering
the effective lagrangian 
\BE
\label{lagrangian}
 {\cal L}_{\rm eff}=  
{{v^2_R}\over{2}}\varphi
       [ c^2_s \Delta - {{\partial^2 } \over{\partial t^2}}] \varphi
      -   T^{\mu}_{\mu} \varphi
\EE
In Eq.(\ref{lagrangian})
higher-order terms in the $\varphi$ -field (or higher derivatives of $\varphi$)
have been neglected 
and the trace of the energy-momentum tensor of ordinary matter is
treated as an external source for $\varphi(x)$. 
This structure holds true
regardless of the spectral decomposition of $\varphi(x)$ for 
${\bf{k}} \to 0$. Of course, 
by choosing $c_s=1$ in Eq.(\ref{lagrangian}) and assuming the wavelengths
of $\varphi(x)$ to cover the full range from zero to infinity,  
the resulting theory
would be exactly Lorentz invariant and correspond to a standard
retarded interaction. 

On the other hand, for the very
large values of $c_s$ that are suggested by the properties of the 
vacuum, the effects of
$\varphi$ have practically no retardation. In this case,
one gets an instantaneous interaction
\BE
\label{step2}
          \Delta \varphi= 
{{ T^{\mu}_{\mu} }\over{c^2_s v^2_R }} 
\EE
of vanishingly small strength when $c_s \to \infty$ (in units of $c=1$).
On the base of Sect.4, we know this limit to correspond to the
continuum theory. 

As discussed by Dicke \cite{varenna}, when averaged over 
sufficiently long times (e.g. with respect to the atomic times), 
by the virial theorem \cite{champ}, the spatial integral of $T^\mu_\mu$ 
represents the total energy of a 
bound system,  i.e. includes the binding energy.
Therefore, for microscopic systems whose components have 
large ${\bf{v}}^2_n/c^2$ but very short periods, this definition 
becomes equivalent to the rest energy. On the other
hand, for macroscopic systems, that have long periods but small 
${\bf{v}}^2_n/c^2$, 
there are no observable differences from the mass density
\BE
\label{rho}
         \rho(x ) = \sum_n M_n
\delta^3( {\bf{x}} - {\bf{x}}_n(t)) 
\EE
and, in the latter case, one re-obtains formally the Poisson equation 
\BE
\label{poisson}
          \Delta U_N 
=G_N \rho(x)
\EE
To this end, the Newton potential $U_N$ is here identified as
\BE
\label{ident}
U_N= \varphi
\EE
and the Newton constant $G_N$ is  expressed as 
\BE
\label{fund3}
         G_N= {{G_F}\over{ c^2_s }}
\EE
If $G_F$ is taken to be the Fermi constant, this gives
 $c_s\sim 4\cdot 10^{16} $ (in units of $c$) as anticipated. 

I observe that the structure of the vacuum suggests
the wavelengths of $\varphi$ to be larger than a centimeter or so.
Therefore, the identification in Eq.(\ref{ident}), with $U_N$ being a
 solution of Eq.(\ref{poisson}), can only hold for suitable 
forms of the mass density, i.e. such that their generated Newton potential 
does not vary appreciably over distances of a centimeter or smaller. 
Experimentally, all known gravitational fields fulfill this condition. 

Notice also that the
Planck mass $M_{\rm Planck}=1/\sqrt{G_N}$ is here
identified with $c_s v_R$ and is not directly 
related to $\Lambda=1/a$. 
For $M_H \sim v_R$, however, one gets 
$c_s v_R\sim \Lambda$ and the two concepts become equivalent. 
In this sense, the Planck scale is not a purely ultraviolet quantity
but embodies in its numerical value the peculiar
infrared-ultraviolet connection that is realized in the scalar condensate.

\section{Geometrizing the phonon dynamics}

{\bf 6.1} After having discussed how the leading effects of 
phonon dynamics can generate Newtonian gravity, one may ask 
whether these phonon effects can be `geometrized', i.e.
re-absorbed into an effective metric structure that depends 
{\it parametrically} on $\varphi(x)$, in agreement 
with the hydrodynamic treatment 
\cite{visser,barcelo,volo2} of superfluid media that exhibits general 
covariance. In this section , I shall outline a
simple answer to this question. My
derivation, however, rather than
a `proof', should be considered a heuristic argument that can be used as
a first approximation. 

As a starting point, let me consider the very peculiar 
space-time picture associated with
the gravitational red-shift. This is usually interpreted as a 
modification of `time'. Namely, the phenomenon is seen as if the
period $T$ of {\it any} clock placed in a gravitational 
potential ($U_N <0$) were
increased, by the precise amount $1-U_N$ 
with respect to the corresponding value $T_o$ of Special Relativity. 
This last quantity can be then defined as the general-relativistic value
 in the limit $U_N \to 0$. 

Now, to accept this interpretation, i.e.
in order to really consider the gravitational red-shift as the effect of a
modification of `time', one should restrict the class of admissible clocks
that can be used to define a measure of time. 
In fact, one should only define a clock through
periodic processes that exhibit the expected slowing down. 
For instance, a pendulum, 
whose period $T_{\rm pendulum} \to \infty$ in the limit of vanishing
gravitational field, would not be an admissible clock. For the same reason
sandglasses, water clocks,...whose existence is solely due to the
gravitational force, should also be discarded. 

Therefore, 
it is reasonable to conclude that the general-relativistic notion of `clock' 
has to be restricted 
to periodic processes whose existence does {\it not} depend
on the gravitational field and for which
there is a well defined zero-gravity limit
\cite{kostro3}. Among these, we find
the atomic clocks, whose period is associated with the spectral lines of
an atomic transition, or their nuclear counterpart, whose period is 
defined in terms of some corresponding
nuclear transition. Actually, due to
their highest precisions, this class of clocks is the {\it only one} 
used so far to detect the tiny effect of a weak gravitational field. 
In any such case,
however, the observed effect on the spectral lines can simply be
understood in terms of the modification of the relevant particle mass (electron or
nucleon) in the given gravitational field. 

To this end, I shall restrict to
ordinary non-relativistic matter (where $T^\mu_\mu$ reduces
to the mass density) and look for 
the effects of a non-zero $\varphi=U_N$ through a
re-definition of any mass placed in the external $U_N$ 
with the tree-level substitution
\BE
\label{tree}
 m_o \to m_o (1+U_N)
\EE
I shall also take into account that 
$U_N$ changes so slowly that its
variation on the atomic scale can be totally neglected. 
In this situation, the effects on the energy levels of 
a hydrogen-like atom simply amounts to a 
re-definition of the electron mass with an average constant value
$m= m_o (1+U_N)$ in  the Dirac Hamiltonian 
\BE
\label{dirac}
        H_D= {\bf{\alpha}}\cdot{\bf{p}} + \beta {m} -{{Ze^2}\over{r}}
\EE
This changes the energy levels and the
frequencies $\omega_o \to \omega_o (1+U_N)$. 
 Therefore, the natural 
period of an atomic clock $T={{2\pi}\over{\omega}}$ is changed, 
$T=T_o(1-U_N)$, with 
respect to the value $T_o={{2\pi}\over{\omega_o}}$ associated with $U_N=0$.
Analogously, the Bohr radius $r_B={{\hbar}\over{ Ze^2 m_o }}$ is changed into
 $ r_B (1-U_N) $ thus producing a symmetric re-scaling of
 the length of the rods. Since all masses are affected in the same way, and
the units of length and time scale as inverse masses, the overall effect 
is equivalent to a conformal 
re-scaling of the metric tensor. To the ${\cal O}(U_N)$ accuracy, we get
\BE
\label{conformal}
                 g_{\mu\nu}(x)=(1-2 U_N)\eta_{\mu\nu}
\EE
where $\eta_{\mu\nu}$ denotes the Minkowski metric.

However, besides the modifications induced on physical rods and clocks, 
the idea of a non-zero
$\varphi=U_N$ to describe  the fluctuations of a (`non-dispersive')
medium, suggests
another physical effect: the introduction of a refractive index in matter-free
space. In this case, 
before the conformal re-scaling, the Minkowski metric would be replaced by
\BE
\label{index0}
\hat{\eta}_{\mu\nu}\equiv({{1}\over{ {\cal N}^2}},-1,-1,-1)
\EE
with a refractive index ${\cal N}={\cal N}(U_N)$ and a normalization
 such that ${\cal N}=1$ when $U_N=0$
(when no confusion can arise, besides the speed of light in the `vacuum' 
$c=1$, I shall also set to unity
the Newton constant $G_N$).

 Thus I obtain the metric structure
\BE
\label{basic01}
         \hat{g}_{\mu \nu}\equiv
({{(1-2 U_N) }\over{{\cal N}^2}},-(1-2U_N),-(1-2U_N),-(1-2U_N))
\EE
that re-absorbs the local, isotropic modifications of Minkowski 
space into its basic ingredients: the value of the speed of light and 
the space-time units. Equivalently, the same metric structure can be 
interpreted as arising from {\it separate}
local changes of the space and time units. 
In fact, such a transformation is known
to represent one of the many possible ways, perhaps the most fundamental, 
to introduce the concept of curvature \cite{feybook,dicke1, szondy}.
In particular, there exist definitions of units, depending
on a scalar field, for which 
a general curved space-time becomes flat, all the Riemannian invariants 
being zero \cite{dicke2}. 

The idea of introducing a refractive index in connection with gravity is
well known (see e.g. \cite{rosen}) and 
very natural, at least when comparing with 
any known medium with definite physical properties. For instance, 
when considering 
a condensate of spinless quanta, and these are treated as hard-spheres, 
Lenz \cite{lenz} first showed that such a system behaves like a medium
with a refractive index. In this approach, the refractive index is not
derived from some dynamical coupling but is
 determined by the geometrical constraints that are placed 
on the propagation of waves of a given wavelength  
 by the presence of the hard spheres. 

Rather than attempting a microscopic 
derivation from first principles, it is much easier to deduce
a possible form of ${\cal N}(U_N)$ using very general arguments. 
To this end, I first observe that, for a time-independent
 $U_N$, the metric Eq.(\ref{basic01}) 
is just a different way to write the general isotropic metric
\BE
\label{isotropic}
 \hat{g}_{\mu \nu} \equiv (A,-B,-B,-B)
\EE
Now, one may ask when the 
local light velocity $c(x,y,z)\equiv \sqrt{ {{ A}\over{B}} }$, 
 defined as a `particle' velocity 
from the condition $ds^2=\hat{g}_{\mu\nu} dx^\mu dx^\nu=0$, agrees with the 
curved-space equivalent of the
phase velocity of light pulses. These are
solutions of  the D'Alembert wave equation with the metric $(A,-B,-B,-B)$
\cite{progress}
\BE
\label{dalembert}
{{1}\over{A}}
{{\partial^2 F }\over{\partial t^2}}
-{{1}\over{B}}(
{{\partial^2}\over{\partial x^2}}+
{{\partial^2}\over{\partial y^2}}+
{{\partial^2}\over{\partial z^2}})F -
{{1}\over{\sqrt {AB^3} }} (\nabla \sqrt{AB})  
\cdot (\nabla F)=0 
\EE
so that, by introducing the 3-vector ${\bf{g}}\equiv 
\sqrt { {{A}\over {B^3}} } (\nabla \sqrt{AB})$ we obtain 
\BE
\label{omega}
{{\partial^2 F }\over{\partial t^2}}
              = {{A}\over{B}}~ \Delta F + {\bf{g}}\cdot (\nabla F)
\EE
By identifying 
${{1}\over{F}}{{\partial^2 F }\over{\partial t^2}}$ as the equivalent 
of $-\omega^2$ and ${{1}\over{F}}\Delta F$ as the corresponding of $-k^2$, 
we find that
particle velocity and phase velocity
$c_{\rm ph}\equiv {{\omega}\over{k}}$ 
agree with each other only when ${\bf{g}}=0$, i.e. when
$AB$ is a constant. This product 
can be fixed to unity with flat-space boundary 
conditions at infinity and, therefore, the resulting value 
\BE
AB=1
\EE
or, in our case
\BE
\label{index}
{\cal N}=(1-2U_N)
\EE
can be considered a consistency requirement on the physical medium, 
if this has 
to preserve, at least to the ${\cal O}(U_N)$ accuracy, the observed
particle-wave duality which is intrinsic in the nature of light. 
On the other hand, if ${\cal N} \neq (1-2U_N)$, one should specify the operative 
definition used for the local speed of light: a) 
the time difference for a light pulse to go forth and back between
two infinitesimally close objects at relative rest, b)
the value obtained combining frequency and wavelength of a given radiation
source,..

With this choice, and to the ${\cal O}(U_N)$ accuracy, 
Eq.(\ref{basic01}) becomes
\BE
\label{basic02}
         \hat{g}_{\mu \nu}\equiv
((1+2 U_N) ,-(1-2U_N),-(1-2U_N),-(1-2U_N))
\EE
thus yielding the first approximation 
to the line element of General Relativity \cite{rosen,einstein} in isotropic
form
\BE
ds^2= (1+ 2U_N) dt^2 - (1-2U_N)(dx^2 +dy^2 +dz^2)
\EE
As anticipated, in this description, the space-time curvature is equivalent to
suitable local re-definitions of the space and time units. Thus, for instance, 
the gravitational red-shift is explained through the behaviour of clocks 
in the gravitational field whereas the energy of the propagating photon 
does {\it not} change with the height \cite{okun} and there is no need to 
introduce the concept of an effective gravitational mass for the propagating 
photons. Analogously, the deflection of light 
can be explained by converting the units of time
and length that define the local speed of light into those that are used 
for the plotted speed \cite{schiff}.  
\vskip 10pt
{\bf 6.2} Let us now try to include the effect of the higher-order 
${\cal O}(U^2_N)$ terms that have been neglected so far by
replacing 
the tree-level relation Eq.(\ref{tree}) with the more general structure
\BE
\label{remi}
 m_o \to  m_o e^{-\sigma}
\EE
where the unknown scalar function $\sigma=\sigma(x)$ in Eq.(\ref{remi})
accounts for the higher powers of $\varphi=U_N$.

 The extension
cannot be done in an universal way. For instance, the 
Schwarzschild metric in isotropic form \cite{weinberg}
\BE
ds^2= {{(1+ U_N/2)^2}\over{(1-U_N/2)^2}}dt^2 - 
(1 -U_N/2)^4 (dx^2 + dy^2 +dz^2)
\EE
violates the condition $AB=1$ to 
${\cal O}(U^2_N)$. On the other hand, insisting on the condition $AB=1$ 
and repeating the same steps as before, 
the metric structure Eq.(\ref{basic02}) is replaced by
\BE
\label{basic2}
         \hat{g}_{\mu \nu}\equiv
(e^{-2\sigma},-e^{2\sigma},-e^{2\sigma},-e^{2\sigma})
\EE
Now, assuming that $\sigma$ can be expanded in powers of $\varphi=U_N$ 
as 
\BE
    \sigma= a_1 U_N + a_2 (U_N)^2 +...
\EE
and comparing Eq.(\ref{basic2}) with Eq.(\ref{basic02}) 
we obtain the following results. 
The first term in
$\sigma$ is just (minus) the Newton potential, i.e. $a_1=-1$. At the same
time, by comparing with all experimental results in a weak centrally 
symmetric gravitational field, one finds $a_2=0$. Therefore, 
up to ${\cal O}({{M^3}\over{r^3}})$ terms, one finds 
\BE 
\label{sigma11}
               \sigma_{\rm exp}= {{M}\over{r}} 
\EE
(or $\sigma_{\rm exp}= {{G_N M}\over{4\pi c^2r}}$ by expressing $M$ in grams and
and $r$ in centimeters). 
If one ignores the cubic and higher-order terms, this
leads to the experimental identification
\BE
\label{exp}
\sigma_{\rm exp}=-U_N
\EE
suggesting that the tree-level redefinition of the particle
mass $m_o \to m_o (1+ U_N+...)$
might actually be turned into its exponentiated form 
$m_o \to m_oe^{U_N}$ after inclusion of 
the higher-order effects. 
Although not totally unexpected, dealing with infrared effects, a full
understanding of
this result requires the all-order evaluation of the 
effective lagrangian for the $\varphi$-field in the presence of the 
external source $T^\mu_\mu$. 

To further appreciate the meaning of Eqs.(\ref{basic2}) and (\ref{sigma11})
let us compare with the 
solutions of Einstein equations. To this end, one can consider
the class of
metrics that are solutions of the following field equations
\BE
\label{general}
 R_{\mu\nu}-{{1}\over{2}}g_{\mu\nu}R
= \gamma (\sigma_\mu \sigma_\nu -{{1}\over{2}} g_{\mu\nu} 
\sigma^\alpha\sigma_\alpha)
\EE
These can be considered Einstein equations in the presence of 
an energy-momentum tensor for a scalar field $\sigma(x)$, 
for various values of the parameter $\gamma$. Indeed, if curvature 
arises from the fluctuations of a medium, it is natural to 
take into account its energy-momentum tensor. Now, for a centrally
symmetric field $\sigma=\sigma(r)$
Tupper \cite{tupper} has shown that {\it all} solutions of Eqs.(\ref{general})
 are 
consistent with the three classical weak-field tests and the
   Shapiro planetary radar reflection experiment. 
Depending on the value of the parameter 
$\gamma$ one gets the Schwarzschild metric, 
for $\gamma=0$, or the Yilmaz metric \cite{yilmaz1}, 
for $\gamma=2$. For any $\gamma$, the solutions 
 are conformal transformations of solutions
of the Brans-Dicke theory \cite{brans}. 

 However, only for $\gamma=2$ the resulting
metric tensor depends {\it parametrically} on $\sigma$. In this case, 
Eqs.(\ref{general}) become algebraic identities
consistently with the point of view that the introduction of the
metric tensor is only an auxiliary tool to effectively account for the
dynamics of the more fundamental field $\sigma(x)$. 
In this case, for $\gamma=2$, where
$\sigma= {{M}\over{r}}$ represents the Newton potential, 
the Yilmaz metric in isotropic form (`Y'=Yilmaz)
\BE
\label{yilma2}
g^{Y}_{\mu\nu}\equiv (e^{-{{2M}\over{r}}} , 
-e^{ {{2M}\over{r}}}, -e^{{{2M}\over{r}}}, -e^{{{2M}\over{r}}})
\EE
is formally identical to Eq.(\ref{basic2}).

In connection with the choice $\gamma=2$, it might be interesting that
the metric Eq.(\ref{basic2}),  differently from all other 
metrics, is not just a one-body solution but is also valid
for a gravitational many-body system where 
 $T_{\mu\nu}\equiv(\rho,0,0,0)$ with 
$\rho$ defined in Eq.(\ref{rho}). 
Indeed, when replacing the Newton potential 
\BE
\label{remark}
         \sigma({\bf{x}} ) =  \sum_n {{M_n}\over
{ |{\bf{x}} - {\bf{x}}_n|}} 
\EE
in the metric Eq.(\ref{basic2}), the field equations 
\BE
\label{general1}
 R_{\mu\nu}-{{1}\over{2}}g_{\mu\nu}R
= 2 (\sigma_\mu \sigma_\nu -{{1}\over{2}} g_{\mu\nu} \sigma^\alpha\sigma_\alpha)
-2 T_{\mu\nu}
\EE
become algebraic identities. In this sense, if one 
wants to compare with real many-body gravitational systems, 
Eq.(\ref{basic2}) represents a very convenient starting point for any
time-dependent approximation. Indeed, for slow motions, the situation
is similar to the conventional adiabatic Born-Oppenheimer 
approximation where one approaches the time-dependent problem
by expanding in the eigenfunctions of a
static 2-center, 3-center,.. n-center hamiltonian.

In connection with a many-body gravitational system, I observe that the 
peculiar factorization properties of the Yilmaz metric, 
$e^{\sum_i {{M_i}\over{r_i}} }=e^{{{M_1}\over{r_1}}} e^{{{M_2}\over{r_2}}}..$,
provide an alternative explanation for the controversial huge 
quasar red-shifts, a large part of which could be interpreted
as being of gravitational (rather than cosmological) origin \cite{clapp}.
 This might represent a
`fifth' test of gravity outside of  the weak-field
regime.

Before concluding, I observe that the metric structure Eq.(\ref{basic2})
with $\sigma=M/r$
was also obtained by Dicke \cite{varenna} in a stimulating remake of 
Lorentz's electromagnetic 
ether. This was based on the simultaneous replacements of the particle mass
\BE
\label{replace1}
                      m_o \to m_o f (\epsilon,\mu)
\EE
and of the light velocity
\BE
\label{replace2}
                      c^2 \to {{c^2}\over{ \epsilon \mu}}
\EE
where $\epsilon$ and $\mu$ are respectively the space-time dependent
dielectric function and 
magnetic permeability of the ether. Consistency with 
the experimental results (the E\"otv\"os experiment, velocity independence of
the electric charge,...) requires $\epsilon =\mu$ leading to the effective
metric structure (`LD'=Lorentz-Dicke)
\BE
\label{lorentz}
 g^{\rm LD}_{\mu\nu}\equiv ({{f^2}\over{\epsilon^4}},
-{{f^2}\over{\epsilon^2}},
-{{f^2}\over{\epsilon^2}},
-{{f^2}\over{\epsilon^2}})
\EE
Finally, a comparison with the classical tests in a weak gravitational field
gives
\BE
\label{ff}
                 f^2= \epsilon ^3
\EE
and 
\BE
\label{ee}
               \epsilon= 1+ 2{{M}\over{r}}+..=e^{2\sigma}
\EE
so that Eq.(\ref{lorentz}) reduces to Eq.(\ref{basic2}) (to compare 
with Eq.(\ref{remi}) one has to rescale $m_oc^2$).

\vskip 10pt
{\bf 6.3} This does not mean
that all possible phenomena that occur in the scalar condensate
can be `geometrized' in terms of Eq.(\ref{basic2}). In fact, 
this metric structure arises naturally in connection with the
longitudinal density fluctuations.
However, not all excitation
states of a superfluid medium, even made up of neutral spinless quanta, 
can be described in terms of a single
scalar function. 

 For instance, 
it is experimentally known \cite{wilks}
that, as a consequence of vortex lines formation, superfluid
 $^4$He can also support circularly polarized transverse waves, 
the so called `Kelvin modes' \cite{tilley}. In this sense, 
vortex line formation in the scalar condensate might represent the key
to obtain genuine graviton states as
quantized transverse oscillations of a string.
Clearly, if such new states 
were substantially excited, there is no reason for 
the associated space-time description to fit with Eq.(\ref{basic2}). 

The connection with string theory can be established within a 
hydrodynamical treatment of the scalar condensate. Independently
of the nature of the elementary constituents of a fluid, 
there is a basic phenomenon in
hydrodynamics: the formation of vortices.
To describe the most general case 
one needs four scalar fields, 
the fluid density $n$ and the three Clebsch potentials
$(\varphi_1, \varphi_2, \varphi_3)$ \cite{lamb,nambu,rasetti,bistro}, 
in terms of which the fluid velocity is expressed as
\BE
     {\bf{u}}= \varphi_1 {\bf { \nabla}} \varphi_2 +{\bf { \nabla}} \varphi_3
\EE
For instance, 
when $\varphi_1=\varphi_2=0$ and $\varphi_3=\varphi$, 
a given configuration of the fluid can be specified through a complex field
\BE
        \psi= \sqrt{n} e^{i\varphi}
\EE
In this way, $|\psi|^2=n$ 
gives the density of the fluid and the phase $\varphi$ defines the fluid 
velocity through ${\bf{u}}={\bf { \nabla}} \varphi$ so that
vortices can only occur in multiply connected regions. Notice that,
in this formalism, although dealing with a complex field, 
there is no fundamental `electric charge' and
all dynamical effects arise from the possible states of
motion of the fluid. 

In this `abelian' case, i.e. 
when the four-dimensional system with vortices 
is modeled as a spontaneously broken U(1) symmetry
\cite{benyacov}, it is known that the resulting theory can be mapped
into a string theory through a duality transformation \cite{leeprd}.
This series of steps, based on the analogy with fluid mechanics, 
leads to the logical
possibility to obtain higher spin states as excitations 
of a simple scalar condensate, i.e. of a medium
whose ultimate constituents are just
 neutral spinless quanta (of size $a\sim 10^{-33}$ cm).

The potential implications of applying the fluid
analogy to the scalar condensate cannot be easily predicted. 
For instance, fluid
mechanics, through the implementation of invariance under 
volume-preserving transformations, provides a simple and elegant 
motivation \cite{jackiw} for the introduction of
 non-commuting gauge fields that are interpreted
in terms of the Lagrange coordinates 
used to label the evolving portions of the fluid. 

As anticipated at the beginning of Sect.5, the problem of transverse 
graviton states
lies outside the scope of this paper. Nevertheless, at least
in the weak-field regime, one expects the resulting classical
space-time picture to agree with General Relativity.
This expectation, quite independently of my proposal, has been  
well illustrated in ref.\cite{barcelo}. According to these authors, 
what we call `Einstein gravity' 
is a kind of universal picture that arises in a large variety of 
systems 
(dielectric media, flowing fluids, Bose condensates, media with a refractive
index, non-linear electrodynamics, thermal vacua,...). 
In this sense, General Relativity (or more precisely
 general covariance) arises in a
coarse-grained description of the world, as hydrodynamics. 
Both, concentrating on 
the properties of matter at scales that are larger than the mean free path for
the elementary constituents, are insensitive to the details of the 
underlying molecular dynamics. 

Of course, even within a general-covariant description, there are some
features, such as the structure of the energy-momentum tensor or the 
presence of a cosmological term $\lambda_c$, 
 that cannot be determined `a priori'  but may depend on the properties 
of the underlying medium.  For instance, as mentioned above 
in connection with the Yilmaz metric, a different energy-momentum tensor
can change qualitatively the space-time description in the strong-field
regime. Analogously, 
a non-zero $\lambda_c$ might modify the causal structure 
of the space-time and give rise to closed time-like loops as solutions
of Einsten equations \cite{deser}. Finally, there can be other effects
if string formation takes place in the scalar condensate.
For instance, by considering
a straight isolated vortex filament along the z-axis, of given mass-density
$\rho$ and given spin-density $J$ per unit length, even 
with a vanishing cosmological term, the resulting geometry supports  
time-like loops \cite{deser}. These can be seen as
 circular paths around the filament
in the $(x,y)$ plane of sufficiently small radius 
\BE
                       r< {{4G_N J}\over{ c^3- 4G_N \rho c}}
\EE

\section{Summary and outlook}

{\bf 7.1} In this paper I have proposed a solution of the `hierarchy problem' 
based on the
 picture of spontaneous symmetry breaking as a real condensation phenomenon
of elementary quanta, the phions \cite{mech}. Skipping the more technical 
details, this physical scenario can be summarized in a series 
of simple basic steps:

~~~i) taking into account
the two-valued nature \cite{legendre,pmu} of the zero-4-momentum
connected propagator in the broken phase, 
 one gets the idea of a true
physical medium that can propagate two types of excitations: a massive one, 
$H(x)$, 
whose energy $E_a({\bf{k}}) \to M_H$ when ${\bf{k}}\to 0$, and that 
corresponds to the usual Higgs boson field,
and a gap-less one, $h(x)$, whose energy
$E_b({\bf{k}}) \to 0$. The latter is naturally interpreted in terms
of the collective density
fluctuations of the system so that 
its wavelengths are larger than $r_{\rm mfp}$, 
the mean free path for the condensed phions. 
 The overall picture is very similar to the 
coexistence of phonons and rotons in superfluid $^4$He that, in fact,
 is usually
considered the condensed-matter analogue of the Higgs condensate.

~~~ii) following the formalism of 
ref.\cite{mech}, the scalar condensate emerges as a highly
hierarchical system characterized by two basic parameters, 
the density $\bar{n}$ and the phion-phion scattering length $a$, 
that, approaching
the continuum limit, combine to produce length 
scales that differ by many orders of 
magnitude. Indeed, $a$, 
the length scale $\xi_H=1/M_H\sim 1/\sqrt{\bar{n} a}$ 
and the phion mean free path 
$r_{\rm mfp}\sim 1/(\bar{n}a^2)\sim 1/\delta$ are related
by 
inverse powers of the diluteness factor $\sqrt{\bar{n}a^3} \to 0$ and
satisfy the relation
\BE
                       \xi^2_H\sim a r_{\rm mfp}
\EE
~~~iii) the mechanism is such that when the ultraviolet cutoff is removed, 
i.e. when
\BE
t={{\xi_H}\over{a}} \sim {{1}\over{\sqrt{\bar{n}a^3} }}
\to \infty
\EE
the acoustic branch $h(x)$ gives rise to infinitesimally weak
(of strength $\epsilon^2={{1}\over{t^2}}$
in units of the Fermi constant) attractive and
nearly instantaneous interactions (with $c_s/c \sim t)$. Therefore, 
the natural interpretation of $h(x)$ is in terms of the Newton potential
$U_N$, thus also providing the underlying rationale for its 
traditional interpretation as an `action at distance'.
As a consequence, my proposal 
might represent a simple physical solution of the
`hierarchy problem' between the Fermi scale $v_R\sim M_H$ 
and the Planck scale $M_{\rm Planck}\sim t v_R$. The latter
is not a purely ultraviolet quantity but embodies in its numerical
value the infrared-ultraviolet connection that is realized in the 
scalar condensate. Numerically one finds 
$M_{\rm Planck}\sim 1/a$ in terms of 
the phion-phion scattering length $a \sim 10^{-33}$ cm. that marks the scale 
up to which phions can be considered as elementary quanta. 

On the other hand, if $h(x)$ were not related to $U_N$, one should,
nevertheless, find some dynamical interpretation. In fact,
rejecting the model presented in this paper without an alternative scenario, 
the existence of $h(x)$ leads to unexplained long-range forces. This
 might represent a serious consistency problem for any physical theory
based on the phenomenon of spontaneous symmetry breaking, such as
the Standard Model.

~~~iv) by further exploiting the scenario where $h(x)=U_N$,
I have suggested that the phonon effects
 can be `geometrized', i.e. re-absorbed into
the space-time metric Eq.(\ref{basic02})
that agrees with General Relativity in weak gravitational fields. 
The extension
to strong fields cannot be done unambiguously and there is more than 
one scenario. For instance, by implementing the particle-wave duality
which is intrinsic in the nature of light, 
a variant of standard General Relativity, Eq.(\ref{basic2}),
 known in the literature as Yilmaz metric \cite{yilmaz1}, might also
be considered. Finally, the analogy with 
superfluid $^4$He suggests the idea of string formation in 
the scalar condensate. In this case, their quantized
transverse excitations might represent a physical model of
genuine graviton states. 

\vskip 10 pt

{\bf 7.2} The comparison with General Relativity requires 
some additional comments.
In the picture presented in this paper, Newtonian gravity 
originates from the long-wavelength
density fluctuations of the scalar condensate. 
These propagate as longitudinal waves, 
starting at wavelengths that are larger than the mean free path
$r_{\rm mfp}$ for the elementary phions and phenomenology requires 
$r_{\rm mfp}$ to be an ${\cal O}(1)$-centimeter length scale.
 As there can be no variation of the 
gravitational potential between two points whose distance is smaller than
$r_{\rm mfp}$, the collective oscillations of the condensate 
average over distances that are much
larger than the atomic size. Thus all quantum 
interference effects disappear. As
these depend on the relative phases of 
the wave functions and, through the particle momentum, on the
particle mass, one understands the origin of
 the so called `weak equivalence principle'. This
states that physical effects in an external 
gravitational field do not depend on the particle mass and, as such, is
at the base of the idea of gravity as
an overall modification of the space-time geometry. 

 An exception is represented, however, 
by those particular experiments where the coherence of the
wave-functions can be maintained over distances where $h(x)$ can 
vary appreciably, as for the 
neutron diffraction experiments in the earth's gravitational field
\cite{cow}. In such cases, the interference pattern depends on the particle
mass. Therefore, the `weak equivalence principle' is no longer valid at the
quantum level. However, for each given type of particle and as a consequence of
the exact equality of inertial and gravitational mass, the interference pattern
obtained in the gravitational field coincide with that obtained in the
equivalent accelerated frame. In this sense, one can conclude that a 
gravitational field is still equivalent to an accelerated frame
(`strong equivalence principle' \cite{cow}).

One more comment is also needed about 
the possible connection with
the traditional induced-gravity approach \cite{fuji,zee,adler}
mentioned in the Introduction. Part of
this discussion was presented in 
ref.\cite{mech}. However, for the convenience of the reader, I'll reproduce
here the basic steps.

The original induced-gravity approach was motivated by the observation 
that for a scalar field $\Phi$
there is a direct coupling  $R \Phi^2$ in the curved-space lagrangian 
density, $R$ being the curvature scalar. Thus, if the field $\Phi$ 
has a non-vanishing expectation value, it was proposed \cite{fuji,zee,adler}
that the Einstein-Hilbert lagrangian could emerge from spontaneous symmetry
breaking, namely 
\BE
\label{hilbert}
          {\cal L}_{Einstein-Hilbert} \sim R \langle \Phi \rangle^2
\EE
Clearly, it goes without saying that the vacuum expectation value in 
Eq.(\ref{hilbert}) cannot be the same $v_R$ related to the
Fermi constant through $G_F\equiv 1/v^2_R$. In fact, in this case, gravity
would have a strength that is $10^{33}$ larger than allowed by 
experiments. Therefore, in the original approach, one was concluding that
the hypothetical
inducing-gravity scalar field could not be the same scalar field 
that induces spontaneous symmetry breaking in the Standard Model.

As discussed in Sect.3, however, 
the non-trivial renormalization effect Eqs.(\ref{phys})-(\ref{zphi}), 
predicted in refs.\cite{zeit,plb,mech} and observed in lattice computations
\cite{cea1,cea2} of the zero-momentum susceptibility, shows that
one is naturally faced with two possible definitions of the
vacuum expectation 
value: a `bare' value $v$ and a physical value $v_R$. They are connected
through their relation to $M_H$, namely
\BE
                M^2_H \sim \lambda v^2 \sim v^2_R
\EE
so that, using Eqs.(\ref{aa}), (\ref{mh}), and (\ref{mfp}) one finds
\BE
     v^2\sim {{M^2_H}\over{a m_\Phi}} \sim {{1 }\over{m_\Phi r_{\rm mfp} }} 
~{{1}\over{a^2}}
\EE
Therefore, within this paper, where $a \sim 10^{-33}$ cm,
the induced-gravity approach can be recovered with two conditions. 
First, 
$\langle \Phi \rangle \equiv v$ is the appropriate value to be used in 
Eq.(\ref{hilbert}). Second, the phion Compton
wavelength has to be of the same order as the phion mean free path, i.e.
\BE
                 m_\Phi \sim {{1}\over{ r_{\rm mfp} }}
\EE
Notice that this condition was found in Sect.4 (see Eq.(\ref{mphirmfp})) on 
the base of the analysis of the scalar condensate. 
Therefore, the present paper can also
provide a convenient framework to recover the original 
induced-gravity approach.

\vskip 10 pt
{\bf 7.3} The mechanism I have proposed
to explain the relative size of the Fermi and Planck
scales, through the introduction of a third infinitesimal momentum
scale $\delta$, 
bears some analogy to other approaches \cite{dimo} based on extra space-time 
dimensions with a `large' compactification size 
$r_c$ (in a regime where
$r_c\sim r_{\rm mfp}\sim 1/\delta$).
In fact, these other approaches, besides solving the `hierarchy problem'
through the same kind of relation 
$L_1/L_2=L_2/L_3$ among three length scales $L_1,L_2,L_3$,
predict, typically, a superluminal 4D effective propagation of gravity 
\cite{csaki}. 

For this reason, independently of my proposal, it becomes more and more
important to have 
an experimental determination of the speed of gravity. However, 
before attempting any experimental determination, 
one should first try to clarify the issue
from a theoretical point of view. For instance, 
can the `speed of Newtonian gravity', $c_N\equiv c_s$, as defined in this
paper through
Eq.(\ref{goldberg}) or Eq.(\ref{lagrangian}), be measured experimentally ?
Or, is there any obvious counterpart in General Relativity ?

Clearly, $c_N$  is 
{\it not} the same concept of the `speed of gravitational waves'. This is 
defined within General Relativity in the weak-field approximation
by expanding the metric tensor around
Minkowski space-time as $g_{\mu \nu}=\eta_{\mu\nu} + h_{\mu\nu}$. 
Further, imposing the transversality condition and
using the formula of quadrupole gravitational radiation, gravitational 
waves are emitted from rapidly varying gravitational
systems that deviate from spherical symmetry. In this framework, 
their propagation
speed has the same value as the speed of light. Their emission, 
when properly taken into account in the energy balance, 
leads to tiny deviations from keplerian orbits that are consistent with 
the observed slowing down of binary pulsars \cite{taylor}.

On the other hand, it is also true that, within the 
presently accepted expanding-universe scenario, 
a cosmological term $\lambda_c$
is now needed to match the experimental observations with
General Relativity. 
 In this case, by using the alternative expansion of the metric tensor
$g_{\mu \nu}= \bar{g}_{\mu\nu} + h_{\mu\nu}$, 
the linearized problem for gravitational waves 
should be cast in the form \cite{curtis}
\BE
              \hat{M}^{\alpha \beta}_{\mu \nu} ( \bar{g}, \lambda_c) 
h_{\alpha \beta}=0 
\EE
where the graviton mass matrix depends on $\lambda_c$ and on the
actual background metric $\bar{g}_{\mu\nu}$ 
used to fit the cosmological data.
Thus, since the sign of the effective
graviton mass squared is not positive-definite \cite{curtis}, 
independently of any experiment to detect gravitational 
waves, the simple idea of massless
gravitons propagating, just as ordinary real photons, at the speed of light is 
not entirely
obvious. For instance, some eigenvalue might correspond to
a complex graviton mass $m_g$.
This problem is usually ignored, probably, because a value
$|m_g|\sim  
\sqrt{|\lambda_c|}$ is so small that the associated `Compton
wavelength' is as large as the visible universe. However, also the
closed time-like loops mentioned in Sect.6 
in connection with a cosmological constant have a typical
radial size $1/\sqrt{|\lambda_c|}$ \cite{deser}. Therefore, the two
problems might be related.

Returning to the speed of Newtonian gravity, 
an experimental value of $c_N$ should be
extracted from the properties of the orbits of gravitationally 
bound systems. 
In this respect, there is no difference between Newtonian gravity
and a static Schwarzschild metric. The well known
differences, such as the precession of perihelia, 
are genuine ${\cal O}(v^2/c^2)$ effects that exist
independently of the speed of gravity. 

Rather, one should study the actual 
motions in the solar system by replacing the absolute-time
description of Newtonian
gravity with a description where
the gravitational influence at $({\bf{r}},t)$ from all possible 
 ${\bf{r}}'$ is evaluated at a retarded
time $t'=t- |{\bf{r}}-{\bf{r}}'|/c_N$. This leads to
modifications in the parameters of the orbits \cite{notetom} that can 
be compared with experiments. 
When this is done, for a large set of data,
Van Flandern finds \cite{tom}
a lower limit $(c_N/c)_{\rm exp}> 2\cdot 10^{10} $. 

Van Flandern's analysis has been criticized \cite{carlip}
on the base of the following observation. 
In ref.\cite{tom} a finite $c_N$ is equivalent to introduce
an `aberration'. Namely, the acceleration felt by a test particle
points towards the retarded position of the gravitational source 
rather than towards its instantaneous position. Within 
Newtonian theory, where gravity is a purely central force, 
 the experimental evidence
for the absence of any gravitational aberration leads to the above lower
limit on $c_N$. However,
it is known from the Lienard-Wiechert potentials of electromagnetism, that
aberration can be cancelled by additional interactions that are proportional
to the velocity of the particle. In this case, despite the potentials are 
retarded (i.e. evaluated at a time $t- r/c$), 
the acceleration, as defined from the Lorentz force, is
always directed along the line that instantaneously connects source
and test particle. Therefore, in principle, introducing
suitable velocity-dependent interactions, the 
absence of aberration can become consistent with a value 
$c_N=c$. This is certainly true and, 
in this sense, any direct 
determination of $c_N$ contains a {\it model dependence} due to
the theoretical framework that is used to compare with experimental data.
Only having a more or less rigid theoretical scenario 
(as in my case, 
where $c_N \equiv c_s$ determines as well
the strength of the central interaction $1/c^2_s r$)
 one can try to obtain an unambiguous answer. 
On the other hand, it is also
true that, in a perspective where
gravity is the excitation of a physical medium, there is no conceptual 
compelling reason for $c_N=c$ (as with the elastic medium considered
in the Introduction for which $c_s \neq c_t$). 

In this context, there are now other attempts to extracts
the speed of gravity from
recent measurements of radio signals past Jupiter in the
experimental conditions of last September 8th, 2002. The idea is
that a finite speed of propagation for 
gravity, affecting the standard Shapiro 
time delay obtained in the static limit, should
 introduce extra effects. As it will be clear in the following, it
is not evident that this other `speed of gravity' is the same
$c_N\equiv c_s$ considered in this paper. Therefore, 
 I'll introduce one more symbol, 
say $c_g$, to indicate the relevant value in this framework. 

To discuss the present situation, it is essential to understand 
how $c_g$ is introduced. The author of ref.\cite{kopeikin} 
starts from the weak-field Einstein equations
\BE
\label{old}
(\Delta - {{1}\over{c^2}} {{\partial^2}\over{\partial t^2}}) 
\gamma^{\alpha\beta}({\bf{r}},t)= 
-{{16\pi G_N}\over{c^4}} T^{\alpha\beta}({\bf{r}},t)
\EE
and argues that to replace 
a value $c_g\neq c$ in the l.h.s. of Eq.(\ref{old}) 
violates Einstein's equations thus
leading to a different theory of gravity. Thus, to remain consistent
with General Relativity, but consider the possibility that
$c_g\neq c$, a re-scaling
of time $t \to \tau$, such that $c_g\tau =ct$, has to be introduced. By
defining $\epsilon=c/c_g$, 
the energy-momentum tensor is then re-scaled as 
$\Theta^{00}=T^{00}$, 
$\Theta^{i0}=\epsilon T^{i0}$, $\Theta^{ij}=\epsilon^2 T^{ij}$ and
Eqs.(\ref{old}) are replaced by 
\BE
\label{new}
(\Delta - {{1}\over{c^2_g}} {{\partial^2}\over{\partial \tau^2}}) 
\gamma^{\alpha\beta}({\bf{r}},\tau)
= -{{16\pi G_N}\over{c^4}} \Theta^{\alpha\beta}({\bf{r}},\tau)
\EE
where finally $c_g$ appears. Within this framework, 
the author of ref.\cite{kopeikin} claims that a measurement of the
extra effects imply
an experimental result $(c_g/c)_{\rm exp}=1.06 \pm 0.21$. 

Another author \cite{asada} concludes, however, that this type of extra
effects is merely due to the time-delay associated
with {\it light} propagation 
(and not to gravity propagation) thus denying that one 
can obtain a measurement of $c_g$. 

Finally, a third author \cite{will}
concludes that "...recent measurements of the propagation of radio
signals past Jupiter are sensitive to $\alpha_1$ but are {\it not} 
directly sensitive to the speed of gravity". 
The same author concludes, however, 
that the existing measurements of the post-newtonian parameter
$|\alpha_1|< 2\cdot 10^{-4}$ should
provide, by themselves, experimental evidence for $c_g=c$. 
To find the relation 
between $\alpha_1$ and $c_g$, the relevant sources are
the papers of ref.\cite{will2}. There the parameter $\alpha_1$ was 
introduced to
take into account deviations from Lorentz-invariance
in the post-newtonian metric through 
a dependence on the velocity ${\bf{w}}$ 
of a given observer with respect 
to the mean rest-frame of the Universe, identified with
the frame of the CMBR. As for velocities
$|{\bf{w}}|\sim 300$ km/sec, no dependence on 
$|{\bf{w}}|$ is observed, 
$|\alpha_1|$ has to be
vanishingly small and gravity has to be 
Lorentz-invariant, i.e. $c_g=c$. 

On the other hand, if gravity were {\it exactly} Lorentz-invariant, 
say as electromagnetism, it is not clear why the dynamics could not be
formulated in ordinary Minkowski space-time. Therefore, 
some genuine Lorentz-non-invariant effect has to be included. The mechanism,
however, does not give rise to observable asymmetries among 
Lorentz-equivalent frames. In this
sense, my proposal introduces the `minimal' departures 
from Lorentz invariance. These
are only due to the vacuum structure while, 
for the rest, the source of $h(x)$ is the Lorentz-invariant 
trace of the energy-momentum tensor in Eq.(\ref{lagrangian}). In such a
framework, there is no way to generate any observable asymmetry.

Finally, following the loose analogy between 
EPR experiments and gravity mentioned
in the Introduction, 
it is interesting that the lower limit for $(c_N/c)_{\rm exp}$ 
quoted by Van Flandern \cite{tom} has a counterpart 
in a similar lower limit for
the speed  of the `quantum information' $c_Q$. 
In fact, for this quantity, an
analysis of long-distance EPR experiments
\cite{scarani} provides the lower limit
$(c_Q/c)_{\rm exp} > 2\cdot 10^{4}$ in the CMBR frame.
\vskip 10pt
{\bf 7.4} Before concluding, I observe 
that, in general, in a superfluid medium one expects
non-linear corrections to the 
free-phonon approximation. As this represents, 
in my picture, the mechanism for Newtonian gravity, I shall try to estimate
the scale where non-linearity shows up by
evaluating a mean free path 
$\zeta_{\rm mfp}$ associated with phonon propagation in the phion condensate.

In superfluid $^4$He, for temperature $T\to 0$, 
the phonon mean free path becomes larger than the size of the container.
Indeed, the typical order of magnitude  relation is  \cite{tisza}
\BE
\zeta_{\rm mfp}(T) \sim 1/f_{\rm norm}(T)
\EE
where $f_{\rm norm}(T)$, the fraction of `normal' fluid in the superfluid
system at a given temperature $T$, is known
to become vanishingly small when $T \to 0$. 
For this reason, the phonon mean free-path 
$\zeta_{\rm mfp} $ is much larger than 
the phion mean free path 
$r_{\rm mfp}\sim {{1}\over{na^2}}$ that we have considered 
so far. 

Actually, $f_{\rm norm}(T=0)$ is non-zero but 
infinitesimally small. In fact, even at zero-temperature, a pure Bose 
condensate cannot be dynamically stable in the presence of 
interactions among the elementary spinless quanta. In the Bogolubov
approximation, these produce a tiny
populations of the lowest 
 $({\bf{k}},-{\bf{k}})$
excited states giving rise
to the depletion Eq.(\ref{deple}). 

In the phion condensate 
where $\sqrt {\bar{n}a^3} = {\cal O}(10^{-16})$, 
this fraction 
$f_D\sim \sqrt{\bar{n}a^3}$ is infinitesimal
and can be taken as a measure of the
residual phonon-phonon
interactions. For this reason, the relevant density 
to determine the phonon mean free path is $f_D\bar{n}$ and not $\bar{n}$. 
Therefore, I would tentatively estimate 
a phonon mean free path 
\BE
\zeta_{\rm mfp} \sim {{1}\over{f_D \bar{n}a^2}} \sim 4\cdot10^{16} r_{\rm mfp}
\EE
to mark the distance over which non-linear effects 
might modify the free-phonon propagation.
Using the previous estimate
$r_{\rm mfp} \sim 1$ centimeter, I find
\BE
\label{refvalue}
\zeta_{\rm mfp} \sim 4\cdot 10^{16}~{\rm cm} \sim 2\cdot 10^3~{\rm AU}
\EE
Using this value, in the relation
for the gravitational acceleration due to the sun
\BE
  {{G_N M_{\rm sun}}\over{ 4\pi \zeta^2_{\rm mfp} }}
\sim 10^{-8}~{\rm cm}\cdot {\rm sec}^{-2}
\EE
one gets a very good agreement with Milgrom's critical acceleration 
value \cite{milgrom} 
\BE
g_o\sim 10^{-8}~{\rm cm}\cdot {\rm sec}^{-2}
\EE
at which one should find 
deviations from a pure Newtonian behaviour. 
In solar-system 
conditions, this type of situation occurs when studying the 
long-term comets, 
those with a period $> 200$ yr and semimajor axis larger than $\sim10^2$ AU. 
To describe some of their
features one has to introduce some `ad hoc' assumptions \cite{tremaine}, and
this might be indicative of deviations from a pure Newtonian behaviour
\cite{comilgrom}.
In this sense, the idea of gravity as a long-wavelength oscillation of 
a superfluid medium leads naturally to the existence of
a non-linear regime. Its precise characteristics, however, are not easy to
predict. 
\vfill
\eject


\begin{thebibliography} {99}
\bibitem{comilgrom}
For instance, in Milgrom's approach, a long-term comet seen around the 
sun is bound when its total energy
$E={{v^2}\over{2}} -{{ G_N M_{\rm sun} }\over{r}}$ is below
a condensation shell $E< \sqrt{G_N M_{\rm sun} g_o}$ 
and not only when $E <0$ \cite{milgort}
\end{thebibliography}
\end{document}